\documentclass[3p,twocolumn]{elsarticle}

\usepackage{makecell}
\usepackage[T1]{fontenc}
\usepackage[english]{babel}		
\usepackage[utf8]{inputenc}
\usepackage{textcomp}
\usepackage{amsmath}
\usepackage{graphicx}
\usepackage{systeme}
\usepackage{multirow}
\usepackage{geometry}
\usepackage{natbib}
\usepackage{adjustbox}
\usepackage[dvipsnames]{xcolor}
\usepackage{mathtools}
\usepackage{subfig}
\usepackage{floatrow}
\usepackage{url}
\usepackage{epstopdf}
\usepackage{url}
\usepackage{subfig}
\usepackage{babel}
\usepackage{amssymb}
\usepackage{etoolbox}
\usepackage{amsthm}
\usepackage[normalem]{ulem}
\usepackage{tikz}
\usetikzlibrary{shapes,arrows}
\usepackage{circuitikz}

\def\@author#1{\g@addto@macro\elsauthors{\normalsize%
		\def\baselinestretch{1}%
		\upshape\authorsep#1\unskip\textsuperscript{%
			\ifx\@fnmark\@empty\else\unskip\sep\@fnmark\let\sep=,\fi
			\ifx\@corref\@empty\else\unskip\sep\@corref\let\sep=,\fi
		}%
		\def\authorsep{\unskip,\space}%
		\global\let\@fnmark\@empty
		\global\let\@corref\@empty  %% Added
		\global\let\sep\@empty}%
	\@eadauthor={#1}
}

\tikzstyle{block} = [draw,fill=white,rectangle, thick,text width=22mm,text centered,minimum height=15mm,font=\bf]
\tikzstyle{sum} = [draw, fill=white,circle,very thick,minimum width=5mm]
\tikzstyle{pinstyle} = [pin edge={to-,very thick,black},font=\bf]

\newcommand{\mytilde}[0]{\raisebox{2.5pt}{\Large\texttildelow}}
\newcommand{\myquote}[1]{``#1''}
\newcommand{\GP}[1]{{\color{black}#1}}
\newcommand{\change}[1]{{\color{black}#1}}
\newcommand{\changeb}[1]{{\color{black}#1}}

\theoremstyle{definition}
\newtheorem{definition}{Definition}

\theoremstyle{remark}
\newtheorem{remark}{Assumption}

\theoremstyle{takeaways}

\begin{document}
\begin{frontmatter}

\title{Exergy-based \change{modeling framework} for hybrid and electric ground vehicles}

\author[a]{Federico Dettù}
\author[a]{Gabriele Pozzato}
\author[b]{Denise M. Rizzo}
\author[a]{Simona Onori\corref{cor1}} 
\address[a]{Energy Resources Engineering Department, Stanford University, Stanford, CA, 94305 USA}
\address[b]{U.S. Army CCDC Ground Vehicle Systems Center, 6501 E. 11 Mile Road, Warren, MI 48397}
\cortext[cor1]{corresponding author}

%\maketitle
\begin{abstract}
Exergy, or availability, is a thermodynamic concept representing the useful work that can be extracted from a system evolving from a given state to a reference \changeb{state}. It is also a system metric, formulated from the first and the second law of thermodynamics, encompassing the interactions between subsystems and the \changeb{resulting} entropy generation. In this paper, an exergy-based analysis for ground vehicles is proposed. The study, a first to the authors' knowledge, defines a comprehensive \changeb{vehicle and powertrain-level} modeling framework to quantify exergy transfer and destruction phenomena \changeb{for the vehicle's longitudinal dynamics and its energy storage and conversion devices (namely, electrochemical energy storage, electric motor, and ICE). To show the capabilities of the proposed model in quantifying, locating, and ranking the sources of exergy losses, two case studies based on an electric vehicle and a parallel hybrid electric vehicle are analyzed considering a real-world driving cycle. This modeling framework can serve as a tool for the future development of ground vehicles management strategies aimed at minimizing exergy losses rather than fuel consumption.}
\end{abstract}

\end{frontmatter}

\section{Introduction}
\changeb{Greenhouse gas emissions reduction and efficiency improvement are fundamental challenges that must be solved to promote sustainability in the transportation sector. Vehicle electrification, in the form of Electric Vehicles (EVs) and Hybrid Electric Vehicles (HEVs), has shown to be the key towards the development of efficient technologies to minimize both powertrain losses and carbon footprint. The design of novel powertrain architectures that meet the above requirements has been traditionally carried out from energy-based analysis. This methodology, however, does not allow for the formal quantification of losses, limiting the further vehicle and powertrain-level optimization.} 

\changeb{On the other hand, exergy, or availability, allows for the quantification of the useful work available to a system and of the associated irreversibilities.} Exergy is an overall system metric that encompasses the interactions between subsystems, and it is formulated upon the first and the second laws of thermodynamics. Exergy is not simply a thermodynamic property of a system, \change{it is also a metric defined with respect to a} reference state: the ability to do work depends upon both the given state of the system and \changeb{its} surroundings. Therefore, the exergy content of a system can change even if the state of the system does not, for instance, if the external temperature \changeb{varies with respect to the initial condition} \cite{camberos_exergy_book}. This fundamental characteristic enables the \change{exergetic analysis of systems interacting with their surroundings.} 

Availability was \change{first} applied to analyze losses in chemical processes and power applications \cite{sciubba_history}. 
Soon after, exergy-based modeling became a powerful tool to assess the overall system performance and help the engineering and designing process in \changeb{other} fields. For instance, applications of availability analysis can be found for gas turbines \cite{Shamoushaki_2017}, \cite{AHMADI20112529}, solar technology \cite{BEHZADI20181011}, \cite{FARAHAT20091169}, nuclear and coal-fired plants \cite{rosen_exergy_nuclear_plant}, \cite{rosen_exergy_coal_nuclear_plants}, and conversion and storage electrical energy systems \cite{rosen_exergy_power_technologies}.

\change{In the aerospace field}, a plethora of work on \change{exergy-based analysis} \changeb{applied to} aircrafts, rockets, and launch vehicles \changeb{can be found}. In particular, \cite{camberos_exergy_book} and \cite{riggins_methodology_performance_analysis} apply \change{exergy} for the design, analysis, and optimization of hypersonic aircrafts. \change{In \cite{watson_launch_vehicles} and \cite{gilbert_exergy_system_engineering}, exergy modeling of rockets and launch vehicles is tackled, respectively. Recent works also show the application of availability to quantify the sources of loss within ships' energy systems. According to \cite{exergy_ship}, this information could be used to further reduce the environmental impact and greenhouse gas emissions in maritime transport.}

Exergy analysis has been widely used \changeb{also} in Internal Combustion Engines (ICEs). The ICE is a complex device, composed of many interacting parts (up to 2000), subject to losses originated from frictions, heat exchange, and suboptimal combustion. In this framework, exergy can be used to optimize combustion and, consequently, braking work generation. Concerning the spark-ignition technology, \cite{RAKOPOULOS20081378} and \cite{valencia2019energy} analyze the exergy transfer and destruction phenomena for synthetic and natural gas fueled engines, respectively. In \cite{rakopoulos_exergy_engine}, an overview on availability modeling of naturally aspirated and turbocharged diesel engines is provided. In \cite{razmara_exergy_engine}, the authors prove that, for a homogeneous charge compression ignition engine, $6.7\%$ of fuel can be saved by implementing an exergy-based Model Predictive Control (MPC) strategy (with respect to a suboptimal reference case). 

\changeb{In the ground vehicle field, researchers exploited exergy analysis at a component-level only and the integration of the different energy storage and conversion devices at the vehicle and powertrain-level has not been addressed yet. A comprehensive exergy-based modeling of the powertrain components and of their interactions and interconnections would allow to classify (for example in terms of thermal exchange, aerodynamic drag, entropy generation in combustion reactions) and quantify inefficiencies, enabling the assessment of how these losses propagate during the vehicle operation.} This information has the potential to enable the development of \changeb{vehicle and powertrain-level} optimization and control strategies aiming at minimizing exergy losses.

The goal of this work is the development of a comprehensive exergy-based modeling framework for ground vehicles, with the ultimate objective of providing a tool for the \changeb{design} of \change{model-based} control and estimation strategies \changeb{based on availability}. \changeb{For the first time, the vehicle's longitudinal dynamics and its energy storage/conversion devices -- electrochemical energy storage, electric motor, and ICE  -- are modeled relying on exergy} \changeb{principles. The proposed framework is control-oriented and modular:} the energy storage and conversion devices are \myquote{building blocks} that can be connected according to the need for vehicle and powertrain-level quantification of availability. \changeb{For a detailed and careful characterization of the exergy state, the thermal behavior of each powertrain component is also considered. In particular, for the electrochemical energy storage device, the thermal model is identified and validated using data collected in our laboratory.} The \change{proposed} framework is applied to two case studies: an EV and a parallel HEV. The analysis, performed in a \textsc{Matlab} simulation environment, allows to locate and quantify the sources of irreversibility \change{within the powertrain}, a \change{key} step to support further optimization of propulsion systems. 

The remainder of the paper is organized as follows. Section \ref{section:theory} summarizes the theoretical concepts related to the second law of thermodynamics and exergy. In Section \ref{section:exergy_balance}, the exergy modeling for the vehicle dynamics, \change{electrochemical energy storage}, electric motor, and ICE is introduced. \change{Then, Section \ref{section:case_study} shows the application of the proposed modeling framework to two case studies and analyzes the simulation results.} Finally, conclusions are \change{outlined} in Section \ref{section:conclusion}.

\section{Exergy Modeling: Theoretical Concepts}\label{section:theory}
In this section, the fundamental concepts related to exergy are introduced. These definitions \changeb{are meant to help} the reader to follow the development of the exergy-based powertrain \changeb{modeling} proposed in Section \ref{section:exergy_balance}. \changeb{The notation, nomenclature, and list of abbreviations are provided at the end of the paper on page 17.}

\changeb{\begin{definition}[Entropy\footnote{\changeb{In accordance with \cite{camberos_exergy_book}, Definition \ref{def:ENT} is based on the Clausius inequality.}}]\label{def:ENT}
Nonconservative quantity representing the inability of a system's energy to be fully converted into work. Considering a heat transfer $Q$ through a boundary surface at temperature $T$, entropy is the state function $S$ satisfying 
\begin{equation}
dS=\frac{dQ}{T}.
\end{equation}
\end{definition}
The second law of thermodynamics and the nonconservative quantity called entropy allow for the explicit quantification of the system irreversibilities.}

\begin{definition}[Exergy]\label{def:EX}
Exergy (or availability) is the maximum useful work (or work potential) that can be obtained from a system at a given state, with respect to a specified \changeb{thermodynamic and chemical reference state}. 
\end{definition}

\begin{definition}[Reference state]\label{def:DS}
\changeb{The reference or dead state (indicated with subscript $0$) is given in terms of its reference temperature $T_0$, reference pressure $\mathcal{P}_0$, and its mixture of chemical species of molar fraction $f_0$. In this state, the useful work (chemical, thermodynamical, mechanical, etc.) is zero and the entropy is at its maximum.}
\end{definition}
\changeb{In this paper, the reference state corresponds to the environment state, i.e., the atmospheric air surrounding the vehicle.}

\begin{definition}[Restricted state]\label{def:RS}
\changeb{The restricted state (indicated with superscript $\star$) refers to a system not at chemical equilibrium with respect to the reference state and where its temperature and pressure are at $T_0$ and $\mathcal{P}_0$, respectively.}
\end{definition}

\begin{definition}[Open and closed systems]
A system is open if it exchanges matter with its surroundings. Conversely, a system \change{who} does not transfer mass with its surroundings is said \change{to be} closed.
\end{definition}

\change{\begin{definition}[Feasible process]
A process is feasible if it \changeb{leads to} positive entropy generation (or, equivalently, exergy destruction).
\end{definition}}

\changeb{The following assumptions are made throughout the paper.}

\begin{remark}\label{remark2}
The reference temperature $T_0$ is lower than or equal to the minimum temperature reached by the system.
\end{remark}

\begin{remark}
The gaseous mixtures considered in this paper are composed by ideal gases only.
\end{remark}

\begin{remark}\label{remark:sv}
\changeb{The system is incompressible, i.e., its volume is constant over time.}
\end{remark}

\subsection{Exergy Balance}\label{section:theory_balances}
\begin{figure}[!t]
	\includegraphics[width=0.75\columnwidth]{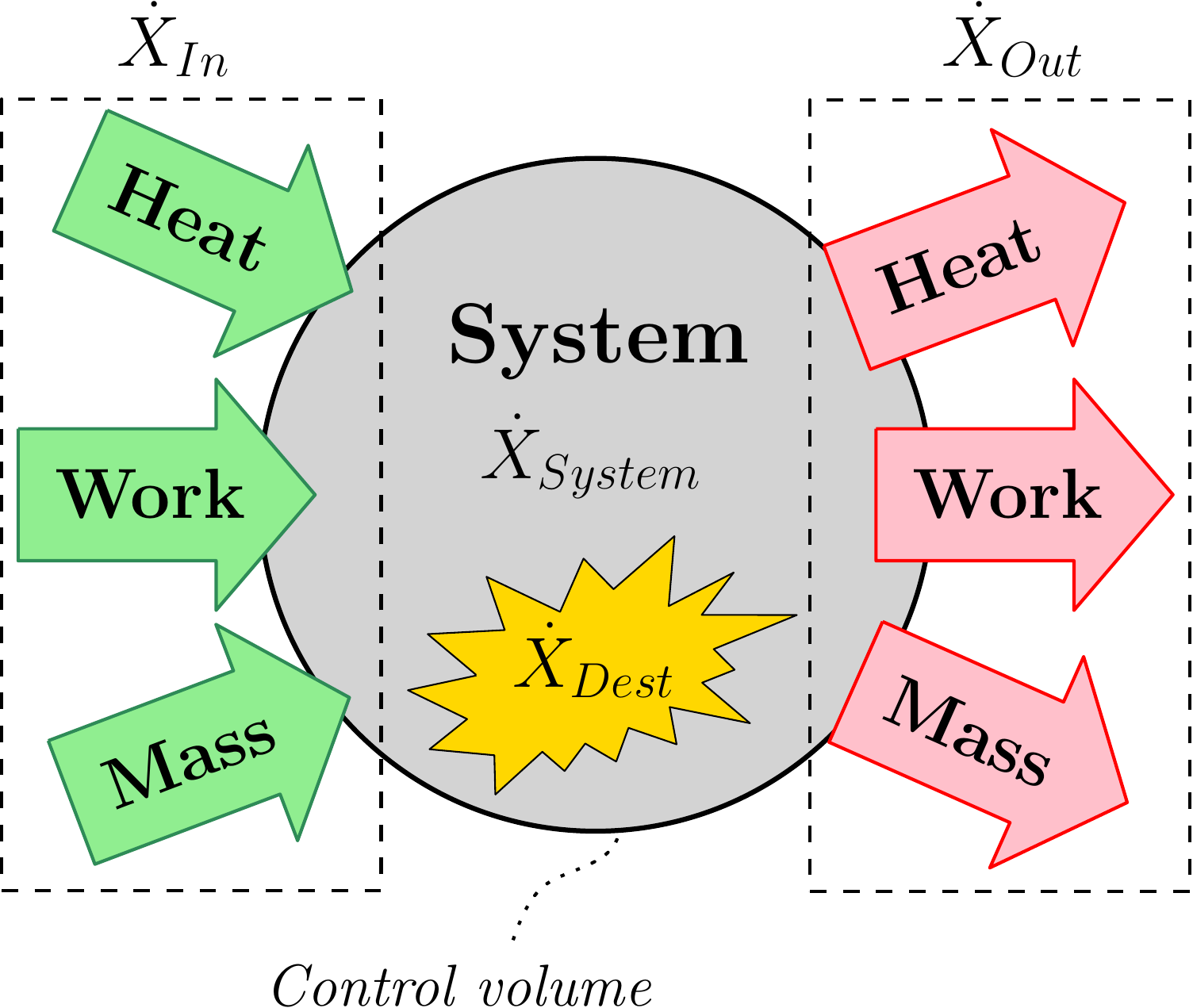}
	\caption{\changeb{Schematic representation of a system exergy balance defined with respect to a control volume. Exergy fluxes -- heat, work, and mass -- entering and leaving the control volume are shown.}}
	\label{fig:exergy_balance_gen}	
\end{figure}

\changeb{The exergy balance of a system is formulated by considering a representative control volume and the heat, work, and mass crossing its boundaries}
\begin{equation}
\dot{X}_{System}=\dot{X}_{In}-\dot{X}_{Out}+\dot{X}_{Dest},
\label{eq:general_balance}
\end{equation}
where $\dot{X}_{System}$ is the exergy of the system, $\dot{X}_{Dest}$ is the exergy \textit{destroyed} due to irreversibilities in the \change{system}, and $\dot{X}_{In}$ and $\dot{X}_{Out}$ are exergy \textit{transfer} terms, modeling \change{heat, work, and mass} fluxes entering \change{and} leaving the control volume, respectively. A pictorial representation of Equation (\ref{eq:general_balance}) is provided in Fig.  \ref{fig:exergy_balance_gen}. 

\subsubsection{Exergy Destruction}\label{section:theory_balances_b}
Any irreversibility in a \change{system}, from frictions to chemical reactions, causes \changeb{entropy increase} and, consequently, exergy destruction. In mathematical terms, this is modeled by the following relationship
\begin{equation}
\dot{X}_{Dest} = -T_0\cdot \dot{S}_{gen},
\end{equation}
where $\dot{S}_{gen}$ is the entropy generation rate. The contribution of exergy destruction is a function of the reference state, which, as already mentioned, must be carefully chosen.

\subsubsection{Exergy Transfer}\label{section:theory_balances_c}
Exergy transfer \change{into} ($\dot{X}_{In}$) or \changeb{out of} ($\dot{X}_{Out}$) the system is \changeb{dependent on} three mechanisms: heat, work, and mass.
\renewcommand{\arraystretch}{1.5}
\begin{table*}[!t]
\begin{center}
\resizebox{1\textwidth}{!}{
\begin{tabular}{|l|c|c|c|c|c|}
\hline
\textbf{Component} & $\boldsymbol{\dot{X}_{System}}$ & $\boldsymbol{\dot{X}_{Heat}}$ & $\boldsymbol{\dot{X}_{Work}}$ & $\boldsymbol{\dot{X}_{Mass}}$ & $\boldsymbol{\dot{X}_{Dest}}$\\ 
\hline\hline
\textit{Longitudinal dynamics (\S~\ref{section:long_dynamics})}& $\dot{X}_{long}$ & $0$ & $P_{trac}$ & $0$ & $-P_{brake}-P_{roll}-P_{aero}$\\ 
\textit{Electrochemical energy storage device (\S~\ref{section:batt_electric_vehicle})} & $\dot{X}_{batt}$ & $\dot{X}_{heat,batt}$ & $\dot{W}_{batt}$ & $0$ & $\dot{X}_{dest,batt}$\\
\textit{\change{Electric motor} (\S~\ref{section:EM_electric_vehicle})} & $\dot{X}_{mot}$ & $\dot{X}_{heat,mot}$ & $\dot{W}_{mot}$ & $0$ & $\dot{X}_{dest,mot}$\\
\textit{ICE (\S~\ref{subsubsection:engine_exergy})} & $0$ & $\dot{X}_{heat,eng}$ & $\dot{X}_{work,eng}+\dot{X}_{intk,eng}$ & $\dot{X}_{fuel,eng}+\dot{X}_{exh,eng}$ & $\dot{X}_{fric,eng}+\dot{X}_{comb,eng}$\\
\hline
\end{tabular}}
\end{center}
\caption{\changeb{Exergy balance terms, defined according to Equation (\ref{eq:overal_ex_theory}), for the vehicle's longitudinal dynamics and its energy storage and conversion devices.}}
\label{tab:exergy_oto}
\end{table*}

\vspace{0.5em}\noindent\textbf{Heat transfer.}  This term is modeled as the maximum possible work extraction from a Carnot heat engine (i.e., an engine operating between two thermal reservoirs \change{at $T_0$ and $T$})
\begin{equation}
\dot{X}_{Heat} = \left(1-\frac{T_0}{T}\right)\cdot\dot{Q},
\label{eq:generic_exheat}
\end{equation}
with $\dot{Q}$ the rate of heat exchange and $T$ the temperature of the \changeb{surface of the system} involved in the thermal exchange. The temperature $T$, at which the thermal exchange occurs, is always higher than the reference state temperature $T_0$ (see \changeb{Assumption} \ref{remark2}). \changeb{Therefore}, $\dot{Q}$ is always negative and leads to a decrease \changeb{of exergy within the system}.

\vspace{0.5em}\noindent\textbf{Work transfer.} This quantity is related to a variation of availability \change{due to} work ($W$) done on or performed by the system. The general expression is 
\begin{equation}
\dot{X}_{Work} = -(\dot{W} - \dot{W}_{Surr})	 = -(\dot{W} - \mathcal{P}_{0}\dot{\mathcal{V}}),
\label{eq:work_ex}
\end{equation}
with $\dot{W}$ the work rate, related to \change{thermomechanical, chemical,} or electrical work, and $\dot{W}_{Surr}$ the moving boundary work, function of the reference state pressure and the system volume time derivative $\dot{\mathcal{V}}$. \change{The term $\dot{W}_{Surr}$ is related to the work the system performs with respect to its surroundings.} \changeb{For instance, during an expansion process, the system volume increases and work is performed to \myquote{move} the surrounding medium.}

\noindent We highlight that there is no entropy transfer associated with work. Instead, being exergy a quantification of the availability, work interactions must be considered in the balance equation: when the system is delivering work ($\dot{W}>0$) the exergy decreases, conversely, if the system is experiencing work ($\dot{W}<0$) the exergy \changeb{increases}.

\vspace{0.5em}\noindent\textbf{Mass transfer.}  \changeb{Exergy can be associated with mass entering or leaving the system. Given a gaseous mixture composed by the chemical species $i$, the variation of exergy with respect to time is expressed as
\begin{equation}
\dot{X}_{Mass}=\sum_i\dot{n_i}\psi_i=\sum_i\dot{n}_i(\psi_{ph,i}+\psi_{ch,i}),
\label{ea:mass_tr_species_i}
\end{equation}
where, for each species $i$, the exergy variation is given by the exergy flux ($\psi_i$) multiplied by the flow rate ($\dot{n}_i$) . In particular, $\psi_i$ is composed of physical $\psi_{ph,i}$ and chemical $\psi_{ch,i}$ exergies. The physical exergy, modeling the work potential between the current state and the restricted state \changeb{of the system}, is expressed as
\begin{equation}
\psi_{ph,i} = h_i - h^{\star}_i-T_0\cdot(s_i-s^\star_i),
\label{eq:phys_ex_general}
\end{equation} 
with $h_i$ and $h^\star_i$ the specific enthalpies of the system at the current state and restricted state, respectively, and  $s_i$ and $s^\star_i$ the specific entropies of the system at the current state and restricted state. To take into account the different chemical composition between the restricted and reference state, the chemical exergy term is expressed as follows
\begin{equation}
\psi_{ch,i} = \mathcal{R}_{gas}\cdot T_0\cdot \log\frac{f^\star_i}{f_{i,0}},
\label{eq:chem_exergy}
\end{equation}
where $\mathcal{R}_{gas}$ is the ideal gas constant, and $f^\star_i$ and $f_{i,0}$ are the molar fractions at the restricted and reference state, respectively. Equation (\ref{eq:chem_exergy}) is valid only for ideal gases. For further details on the exergy flux, the reader is referred to \cite{rakopoulos2009diesel}.}

\subsubsection{Open and Closed Systems}
From the concepts introduced in Sections \ref{section:theory_balances_b} and \ref{section:theory_balances_c}, \changeb{the rate of exergy change for an open system is defined as}
\begin{equation}
\begin{split}
\dot{X}_{System} =& \overbrace{\left(1-\frac{T_0}{T}\right)\cdot\dot{Q}}^{\text{Heat transfer }\dot{X}_{Heat}}\ \overbrace{-(\dot{W}-\dot{W}_{Surr})}^{\text{Work transfer }\dot{X}_{Work}}+\\
                                &+\underbrace{\sum_j\dot{n}_j\psi_j-\sum_k\dot{n}_k\psi_k}_{\text{Mass transfer }\dot{X}_{Mass}}\ \underbrace{-T_0\cdot \dot{S}_{gen}}_{\text{Destruction }\dot{X}_{Dest}}.
\end{split}
\label{eq:overal_ex_theory}
\end{equation}
\changeb{The exergy associated with the mass transfer is defined starting from Equation (\ref{ea:mass_tr_species_i}), considering both the species entering ($i=j$) and exiting ($i=k$) the control volume. The species moving into the control volume lead to an exergy increase. Conversely, the species exiting the control volume lead to an exergy decrease. From Assumption \ref{remark:sv} and Equation (\ref{eq:work_ex}), $\dot{W}_{Surr}$ is equal to zero.

As a reference for the next sections, in Table \ref{tab:exergy_oto}, the exergy balance in Equation (\ref{eq:overal_ex_theory}) is defined for the vehicle's longitudinal dynamics and its energy storage and conversion devices.}

\section{Vehicle Model and Exergy Balance}
\label{section:exergy_balance}
In this section, the modeling of the vehicle longitudinal dynamics, electrochemical energy storage device, electric motor, and ICE is introduced and the exergy balance is \changeb{carried out} relying on the theoretical concepts presented in Section \ref{section:theory} \changeb{and, in particular, in Table \ref{tab:exergy_oto}.} These components are \myquote{building blocks} to be used \changeb{in the description of the overall vehicle architecture}, as shown in Section \ref{section:case_study}.

\subsection{Reference State}
\label{section:ref_env}
In this work, the reference state, as defined in Definition \ref{def:DS}, is at temperature $T_0=298.15K$ and pressure $\mathcal{P}_{0}=1atm$. The atmosphere is assumed to be exclusively composed by \change{four species $k\in\mathcal{K}=\{N_2,O_2,H_2O,CO_2\}$}, and other components \change{(mostly argon)}, combined according to the following molar fractions \cite{mahabadipour_engine}
\begin{equation}
\begin{split}
	&f_{N_2,0}=0.7567,\ f_{O_2,0}=0.2035,\ f_{CO_2,0}=0.0003,\\ &f_{H_2O,0}=0.0303,\ f_{others,0}=0.0092.
\end{split}
\end{equation}

\subsection{Vehicle Model}
\changeb{The exergy-based modeling framework relates the vehicle's longitudinal dynamics, the electrochemical storage device, electric motor, and ICE (when used).} The transmission is assumed to have \changeb{zero} losses and to transfer all the mechanical power from and to the powertrain. This is in line with \cite{hu2013energy}, in which transmission efficiencies close to 100\% are reported.
 
\subsubsection{Longitudinal Dynamics}
\label{section:long_dynamics}
\changeb{A} vehicle model is based on its longitudinal dynamics\footnote{Since we are not interested in the detailed behavior of the vehicle's sprung mass (vertical dynamics) or of the  tires (lateral dynamics), this is a reasonable assumption.}. \change{Without loss of generality}, in this work the road grade component is neglected. \changeb{To simplify the notation,} time dependency is made explicit only the first time a variable is introduced.

The following balance of forces governs the longitudinal dynamics
\begin{equation}
m_{veh}\cdot \dot{v}(t)=F_{trac}(t)-F_{brake}(t)-F_{roll}(t)-F_{aero}(t),
\label{eq:long_dyn}
\end{equation}
where $v(t)$ and $\dot{v}$ are the vehicle speed and acceleration, respectively, $m_{veh}$ the vehicle mass, $F_{trac}$ is the traction force at the wheels, and $F_{brake}$, $F_{roll}$, $F_{aero}$ are the braking, rolling friction, and aerodynamic drag forces experienced by the vehicle, respectively. The forces are computed according to the following equations \change{\cite{simona_book}}
\begin{equation}
	\begin{split}
	F_{aero}&=\dfrac{1}{2}A_f\cdot \rho_{air}\cdot C_d \cdot v^2,\\
	F_{roll}&=m_{veh}\cdot g\cdot k_{roll},\\
	F_{brake}&=\dfrac{\tau_{brake}}{\mathcal{R}_{wh}},\\
	F_{trac}&=\dfrac{P_{trac}(t)}{v(t)},
	\end{split}
\end{equation}
where $P_{trac}$ is the traction power, \change{$\tau_{brake}$} is the braking torque at the wheels, $k_{roll}$ is the rolling friction coefficient, $A_f$ is the vehicle frontal area, $C_d$ is the aerodynamic drag coefficient, and $\rho_{air}$ is the air density. Note that $P_{trac}$, and consequently $F_{trac}$, is a function of the powertrain architecture. Multiplying both sides of Equation (\ref{eq:long_dyn}) by $v$, the vehicle power balance at the wheels is obtained
\begin{equation}
\begin{aligned}
	P_{long}(t)&=m_{veh}\cdot \dot{v}\cdot v
	=\left(F_{trac}-F_{brake}-\right. \\
	&\left.-F_{roll}-F_{aero}\right)\cdot v=P_{trac}(t)-P_{brake}(t)-\\
	&-P_{roll}(t)-P_{aero}(t),
\end{aligned}\label{eq:lewis_hamiltonian}
\end{equation}
where $P_{brake}$, $P_{roll}$, $P_{aero}$ are the powers associated to $F_{brake}$, $F_{roll}$, and $F_{aero}$, respectively. \changeb{Recalling that the Hamiltonian of a system is the sum of kinetic and potential energy, Equation (\ref{eq:lewis_hamiltonian}) can be interpreted as the derivative of the Hamiltonian, i.e., $P_{long} = \dot{H}(t)$. In this particular case, the road grade is assumed to be zero and the potential energy contribution to the Hamiltonian is zero.} According to \cite{robinett_book}, under the assumptions
\begin{itemize}
\item no heat flow,
\item no exergy flow,
\item no mass flow rate,
\end{itemize}
the following equality holds
\begin{equation}
\dot{X}_{long}(t)= \dot{H} = P_{long},
\label{eq:hamiltonian_exergy}
\end{equation}
i.e., the derivative of the Hamiltonian \changeb{function} of the system is equal to its exergy rate, indicated with $\dot{X}_{long}$. 

For the electrochemical energy storage device, electric motor, and ICE \changeb{(used in the HEV)}, the assumptions listed above do not hold because of heat exchanges with the environment, work, entropy generation, and mass transfer. \changeb{For these components, the exergy rate balance is not based on the sum of kinetic and potential energy only, thus, it is not equal to the derivative of the Hamiltonian.} In the next sections, a careful formulation of the exergy balance for these powertrain components is carried out.
\begin{figure}[!b]
	\begin{center}
		\begin{circuitikz}
			\draw (0,0)
			to[V,v=$V_{oc}^{cell}$] (0,2)
			to[R=$R_0^{cell}$,-*,i^>=$I_{cell}$] (2,2)
			to[open,v^<=$V_{cell}$] (2,0)
			to[short,*-] (0,0);				
		\end{circuitikz}
	\end{center}
	\caption{\changeb{Schematic of the battery cell zero-order model.}}
	\label{fig:bat_pack}
\end{figure}
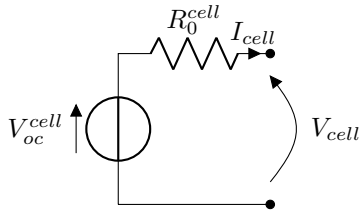

\subsubsection{Electrochemical Energy Storage Device}
\label{section:batt_electric_vehicle}
An \change{electrochemical energy storage device}, in the form of a lithium-ion battery pack, is used in both EV and HEV powertrains. \changeb{The model is first derived at cell-level and then upscaled to the pack-level.}

\begin{figure}[!tb]
\centering
\includegraphics[width=\columnwidth]{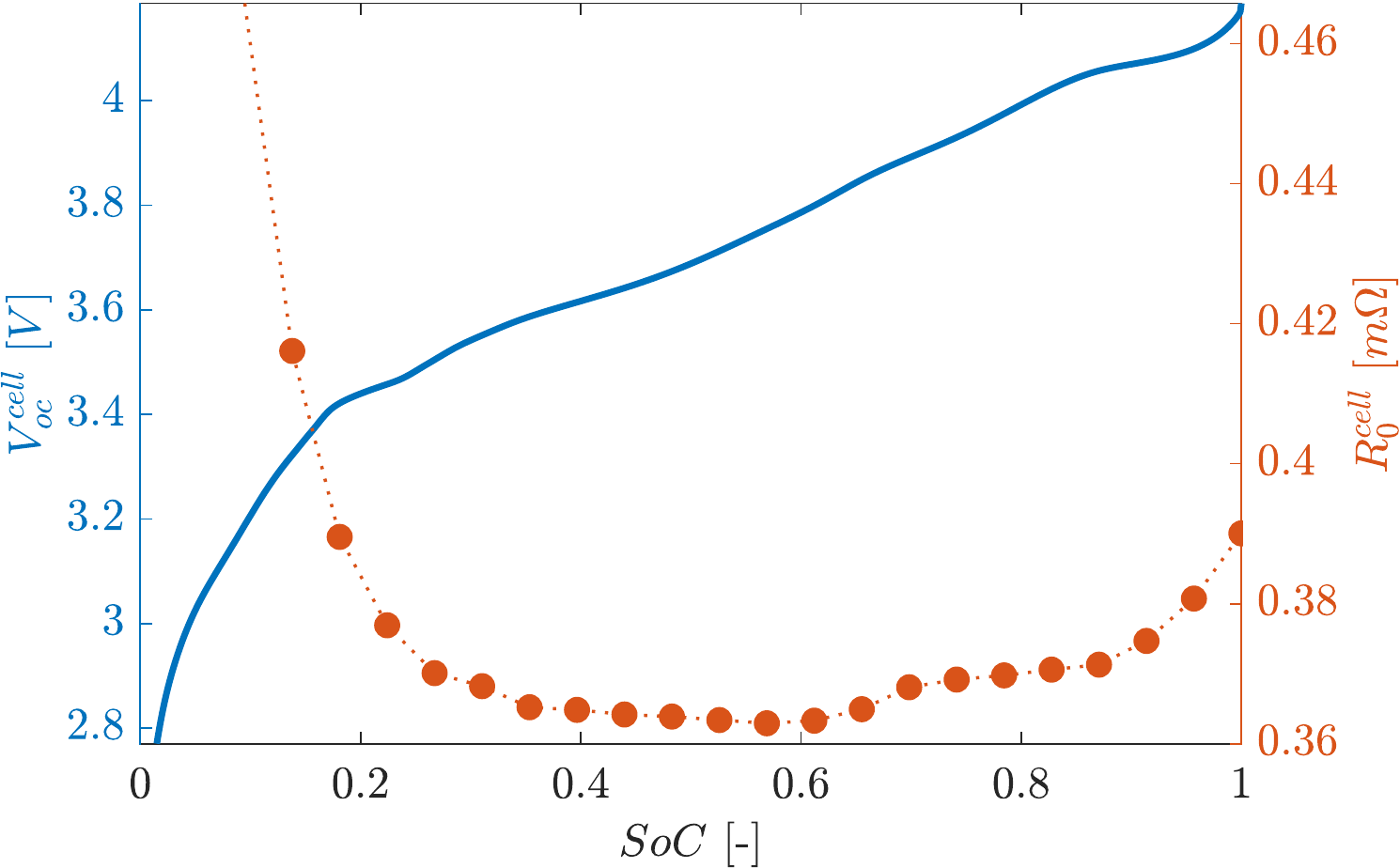}
\caption{Battery cell open circuit voltage and internal resistance \changeb{as a function of $SoC$} at $298K$.}\label{fig:soc_voc_r0}
\end{figure}

\begin{figure}[!tb]
	\centering
	\includegraphics[width=\columnwidth]{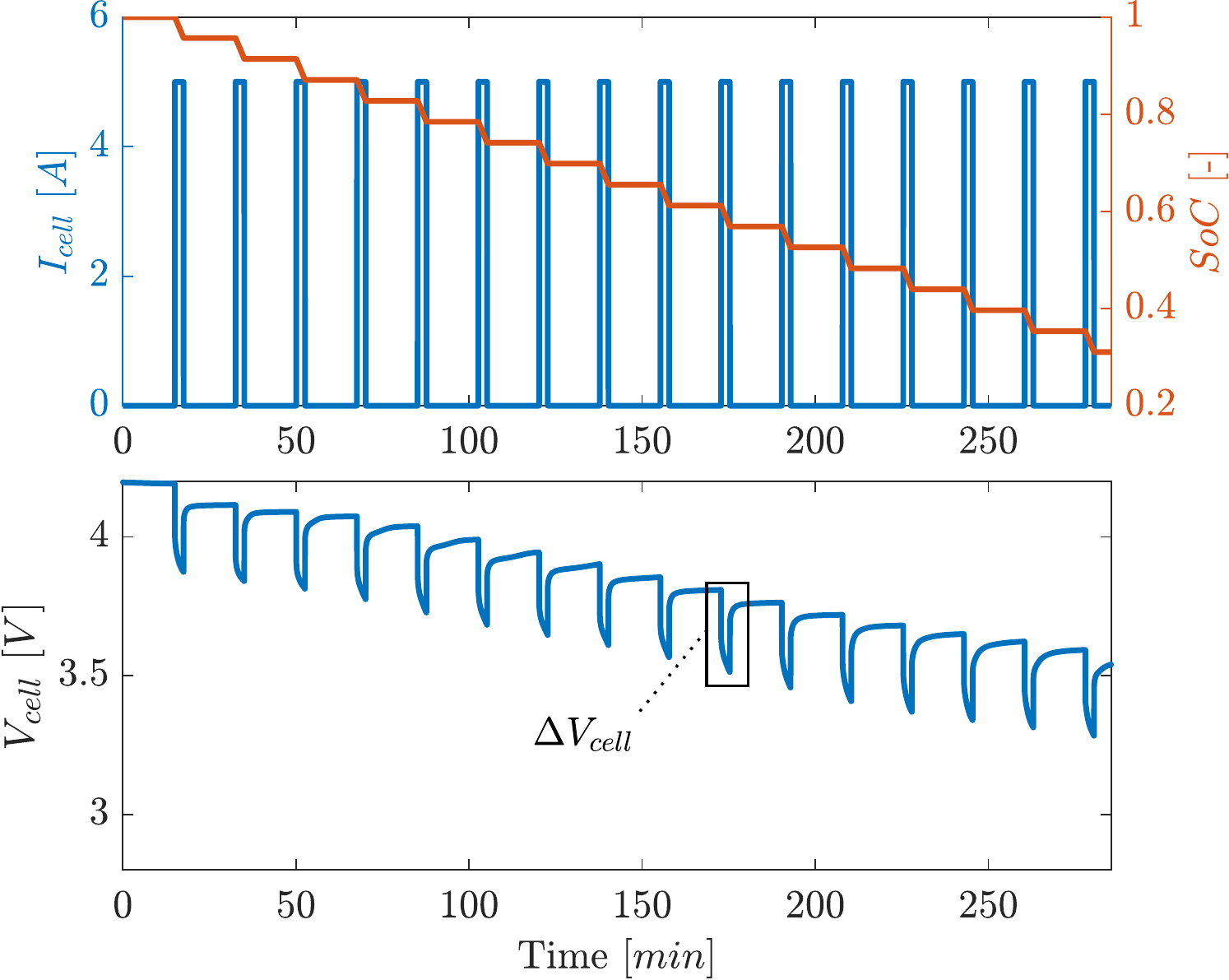}
	\caption{\changeb{Current pulse train discharge test and corresponding $SoC$ and terminal voltage $V_{cell}$. The voltage drop $\Delta V_{cell}$ at 0.6 $SoC$ is highlighted.}}\label{fig:battery_ele}
\end{figure}

\changeb{For the purpose of modeling the losses in the battery cell, a zero-order equivalent circuit model, as the one shown in Fig. \ref{fig:bat_pack}, is used. From Kirchhoff Voltage Law, the terminal voltage is given by
\begin{equation}
	V_{cell}(t)=V_{oc}^{cell}(t)-R_0^{cell}\cdot I_{cell}(t),
	\label{eq:kvl}
\end{equation}
where $V_{oc}^{cell}$ is the cell open circuit voltage, $R_0^{cell}$ is the lumped internal resistance, and $I_{cell}$ is the battery cell current. The convention for $I_{cell}$ is as follows
\begin{equation}
\begin{split}
&\begin{array}{c}
\begin{cases}
I_{cell}\leq 0 & \textrm{if the battery is charging},\\
I_{cell}> 0 & \textrm{otherwise}.
\end{cases}\end{array}
\end{split}
\end{equation}
The battery State of Charge ($SoC$) dynamics is defined as
\begin{equation}
	\dot{SoC}(t)=-\frac{I_{cell}}{3600\cdot Q_{nom}^{cell}},
	\label{eqsoc}
\end{equation}
where $Q_{nom}^{cell}$ is the battery cell nominal capacity in $Ah$. From the cell nominal voltage $V_{nom}^{cell}$, the device nominal energy is obtained as
\begin{equation}
	E_{nom}^{cell}=V_{nom}^{cell}\cdot Q_{nom}^{cell}\cdot 3600.
\end{equation}
\changeb{From} the battery cell power $P_{cell}(t)=V_{cell}\cdot I_{cell}$, the State of Energy ($SoE$) dynamics is defined as
\begin{equation}
\dot{SoE}(t)=-\dfrac{P_{cell}}{E_{nom}^{cell}}.
	\label{eqsoe}
\end{equation}
The cell internal resistance $R_0^{cell}$ and the open circuit voltage $V_{oc}^{cell}$ as a function of $SoC$ are shown in Fig. \ref{fig:soc_voc_r0}. These data have been collected at the Stanford Energy Control Lab from a LG Chem INR21700-M50 NMC cylindrical cell with nominal capacity $Q_{nom}^{cell} = 4.85Ah$. The internal resistance is identified according to the procedure described in \cite{catenaro2021experimental} with data  from \cite{catenaro2021dib}, namely, the cell is discharged through the current pulse train shown in Fig. \ref{fig:battery_ele}. Then, the voltage drop $\Delta V_{cell}$, after each pulse, is evaluated and divided by the measured current $I_{cell}$ to obtain the cell internal resistance at different $SoC$.}

\begin{figure}[!tb]
	\centering
	\includegraphics[width=\columnwidth]{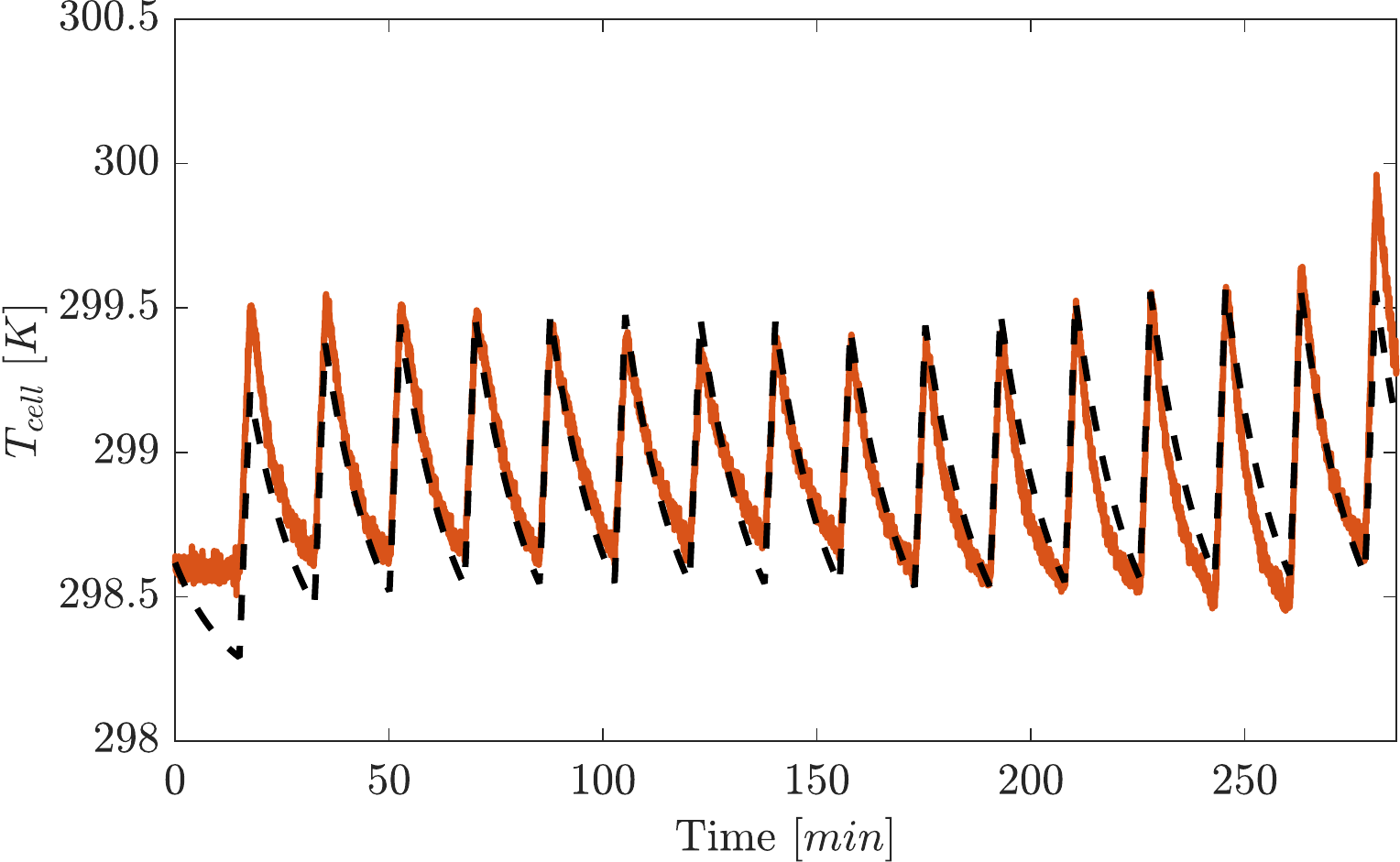}
	\caption{\changeb{Comparison between the experimental and simulated temperature profile during the current pulse train discharge test of Fig. \ref{fig:battery_ele}.}}\label{fig:battery_temperature}
\end{figure}

\changeb{The battery cell temperature $T_{cell}(t)$ and heat transfer with the environment play a relevant role in the evaluation of the exergy balance shown later. Thus, the following lumped thermal model is introduced
\begin{equation}
	\mathrm{C}_{cell} \cdot\dot{T}_{cell}=(V_{oc}^{cell}-V_{cell})\cdot I_{cell}+\dot{Q}_{cell}(t),
	\label{eq:battery_thermalcell}
\end{equation}
where $\mathrm{C}_{cell}$ is the battery cell thermal capacity, the first term on the right-hand-side is the heat generation due to Joule losses, and $\dot{Q}_{cell}$ is the heat transfer between the device and the environment, defined as
\begin{equation}
	\dot{Q}_{cell}=\mathrm{h}_{out,cell}\cdot (T_0-T_{cell}),
	\label{eq:battery_exchangecell}
\end{equation}
with $\mathrm{h}_{out,cell}$ the thermal transfer coefficient between the battery cell and the environment, at the reference temperature $T_0$. The cell thermal capacity ($\mathrm{C}_{cell}$) and thermal transfer coefficient ($\mathrm{h}_{out,cell}$) are identified using the current pulse train discharge test shown in Fig. \ref{fig:battery_ele}. In particular, the identification is performed minimizing the Root Mean Square Error (RMSE) between the cell experimental temperature (measured by a thermocouple) and the simulated one. Fig. \ref{fig:battery_temperature} shows a comparison between measured and simulated temperature profiles. In this context, a RMSE of $0.114K$ is obtained.}

\changeb{From the cell-level modeling, electrical and thermal quantities are upscaled to the pack-level as 
\begin{equation}
\begin{split}
& R_0\left(SoC\right) = \frac{N_s}{N_p}R_0^{cell}\left(SoC\right),\\
& V_{oc}\left(SoC\right) = N_s\cdot V_{oc}^{cell}\left(SoC\right),\\
& V_{nom} = N_s\cdot V_{nom}^{cell},\\
& Q_{nom} = N_p\cdot Q_{nom}^{cell},\\
& E_{nom}=V_{nom}\cdot Q_{nom}\cdot 3600,\\
& \mathrm{C}_{batt} = N_s\cdot N_p\cdot\mathrm{C}_{cell},\\
& \mathrm{h}_{out,batt} = N_s\cdot N_p\cdot\mathrm{h}_{out,cell},
\end{split}
\label{eq:upscaled}
\end{equation}
with $N_s$ and $N_p$ the number of cells in series and parallel configuration. Given the parameters in Equation (\ref{eq:upscaled}), Equations (\ref{eq:kvl}), (\ref{eqsoc}), (\ref{eqsoe}), (\ref{eq:battery_thermalcell}), and (\ref{eq:battery_exchangecell}) are rewritten as
\begin{equation}
\begin{split}
&V_{batt}(t)=V_{oc}(t)-R_0\cdot I_{batt}(t),\\
&\dot{SoC}=-\frac{I_{batt}}{3600\cdot Q_{nom}},\\
&\dot{SoE}= -\dfrac{P_{batt}(t)}{E_{nom}} = -\dfrac{V_{batt}\cdot I_{batt}}{E_{nom}},\\
&\mathrm{C}_{batt} \cdot\dot{T}_{batt}(t)=\ (V_{oc}-V_{batt})\cdot I_{batt}+\dot{Q}_{batt}(t),\\
& \dot{Q}_{batt}=\mathrm{h}_{out,batt}\cdot (T_0-T_{batt}).
\end{split}
\label{eq:upscaledall}
\end{equation}}

The battery pack is assumed to be a closed system, as no mass is exchanged between the device and the environment. \changeb{Thus, according to \cite{camberos_exergy_book}, Equation (\ref{eq:overal_ex_theory}) can be written as}
\begin{equation}
	\dot{X}_{batt}(t)=\dot{X}_{heat,batt}(t)+\dot{W}_{batt}(t)+\dot{X}_{dest,batt}(t),
	\label{eq:battery_exergy}
\end{equation}
where $\dot{X}_{batt}$ is the battery exergy rate, $\dot{W}_{batt}=-P_{batt}$ is the work rate to/from the battery and $\dot{X}_{dest,batt}=-T_0\dot{S}_{gen,batt}(t)$ is the exergy destruction within the battery, where $\dot{S}_{gen,batt}$ is in turn the entropy generation rate. $\dot{X}_{heat,batt}$ is the exergy transfer contribution due to heat transfer computed as in Equation (\ref{eq:generic_exheat})
\begin{equation}
	\dot{X}_{heat,batt}=\left(1-\dfrac{T_0}{T_{batt}}\right)\cdot \dot{Q}_{batt}.
	\label{eq:batt_heat_ex}
\end{equation}
\change{The entropy generation rate is computed formulating the entropy balance for the battery ($\dot{S}_{batt}$), while recalling the closed system assumption \cite{camberos_exergy_book}. In this scenario, the entropy variation is due to the heat transferred to or from the system and the entropy generation}
\begin{equation}
	\dot{S}_{batt}(t)=\dot{S}_{gen,batt}+\dot{S}_{in,batt}(t)-\dot{S}_{out,batt}(t).
	\label{eq:entropy_rate_battery}
\end{equation}
Given that $T_{batt}\geq T_0,\ \forall t$, then $\dot{S}_{in,batt}=0$ (i.e., the entropy transfer rate from the environment to the battery is zero), since the heat is always going from the device to the environment. On the other hand, the entropy transfer rate from the battery to the environment is obtained as
\begin{equation}
	-\dot{S}_{out,heat}=\dfrac{\dot{Q}_{batt}}{T_{batt}}=\dfrac{\mathrm{h}_{out,batt}\cdot \left(T_0-T_{batt}\right)}{T_{batt}}.
	\label{eq:entropy_out_battery}
\end{equation}
According to \cite{doty_entropy_unsteady}, the entropy for a closed system is a function of its states
\begin{equation}
S_{batt}=S_{batt}\left(T_{batt},\mathcal{P}_{batt}(t)\right),
\label{eq:entropy_batt}
\end{equation}
where $\mathcal{P}_{batt}$ is the battery pressure. Recalling that the system is incompressible \changeb{(Assumption \ref{remark:sv})}, \changeb{the pressure dependence of Equation (\ref{eq:entropy_batt}) can be removed}, leading to entropy generation and transfer due to heat exchange only. \changeb{Given Definition \ref{def:ENT}, the following relationship hold
\begin{equation}
\begin{aligned}
	&dS_{batt}=\dfrac{\mathrm{C}_{batt}\cdot dT_{batt}}{T_{batt}},
\label{eq:battery_entropy_1}
\end{aligned}
\end{equation}
and, dividing both sides of Equation (\ref{eq:battery_entropy_1}) by $dt$, \changeb{Equation} (\ref{eq:entropy_rate_battery}) is rewritten as}
\begin{equation}
	\dot{S}_{batt}=\dfrac{\mathrm{C}_{batt}}{T_{batt}}\cdot \dot{T}_{batt}.
	\label{eq:battery_entropy_2}
\end{equation}
Recalling Equations (\ref{eq:upscaledall}), (\ref{eq:entropy_rate_battery}), (\ref{eq:entropy_out_battery}), and (\ref{eq:battery_entropy_1}), the final expression for the entropy generation rate is retrieved
\changeb{\begin{equation}
	\dot{S}_{gen,batt}=\dfrac{\left(V_{oc}-V_{batt}\right)\cdot I_{batt}}{T_{batt}} = \frac{R_0\cdot I_{batt}^2}{T_{batt}},
	\label{eq:battery_entropy_rate_gen}
\end{equation}}
which is always positive, coherently with the second law of thermodynamics. 

Finally, the battery exergy balance is obtained substituting Equations (\ref{eq:batt_heat_ex}) and (\ref{eq:battery_entropy_rate_gen}) into Equation (\ref{eq:battery_exergy})
\begin{equation}
\begin{split}
\dot{X}_{batt} &= \left(1-\frac{T_0}{T_{batt}}\right)\cdot\dot{Q}_{batt}-P_{batt}-\\
&\hspace{1em}-\frac{T_0}{T_{batt}}R_0\cdot I_{batt}^2.
\end{split}
\label{eq:battery_fullex}
\end{equation}

\subsubsection{Electric Motor}
\label{section:EM_electric_vehicle}
\begin{figure}[!b]
	\centering
	\subfloat[Electric Vehicle.]{\includegraphics[width=\columnwidth]{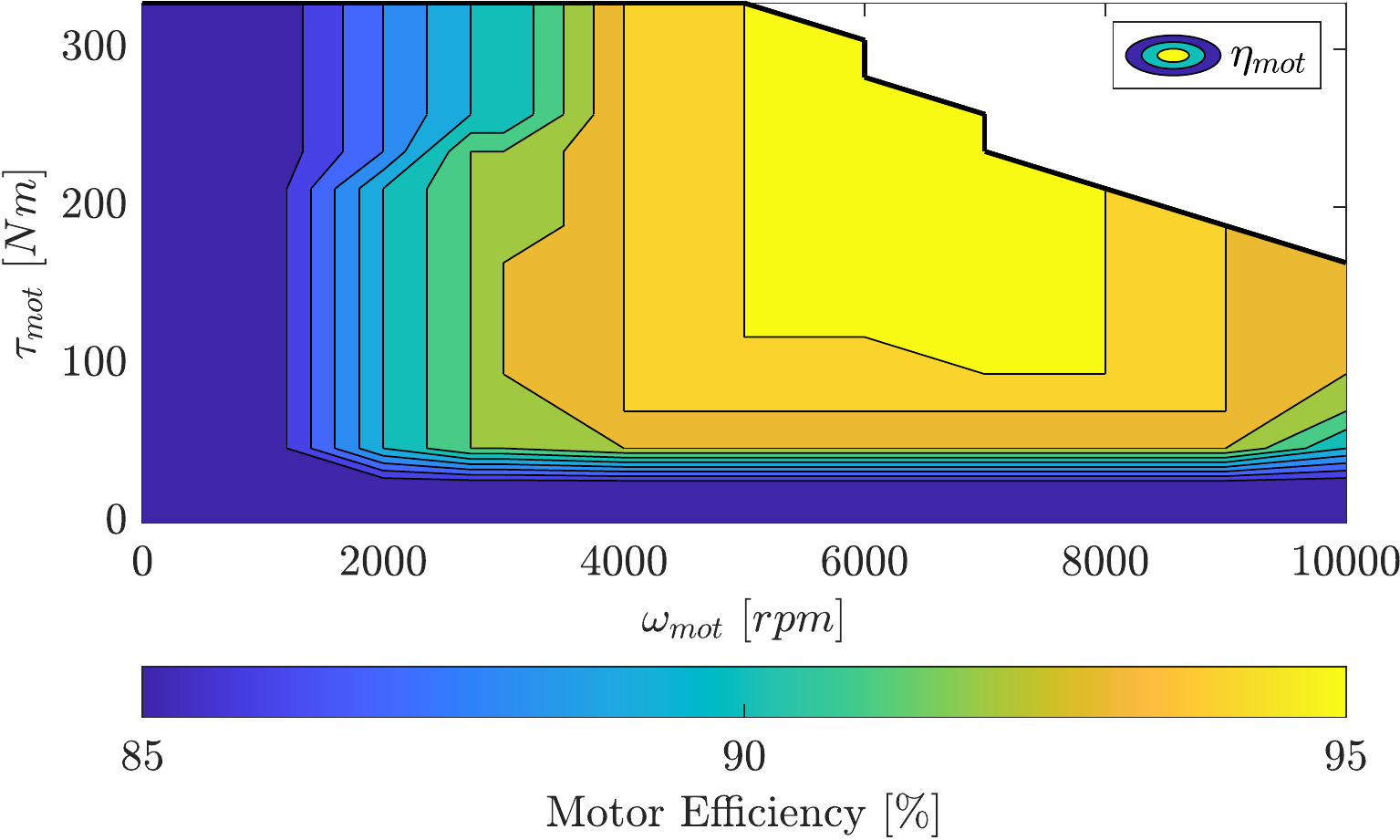}}\\
	\subfloat[Hybrid Electric Vehicle.]{\includegraphics[width=\columnwidth]{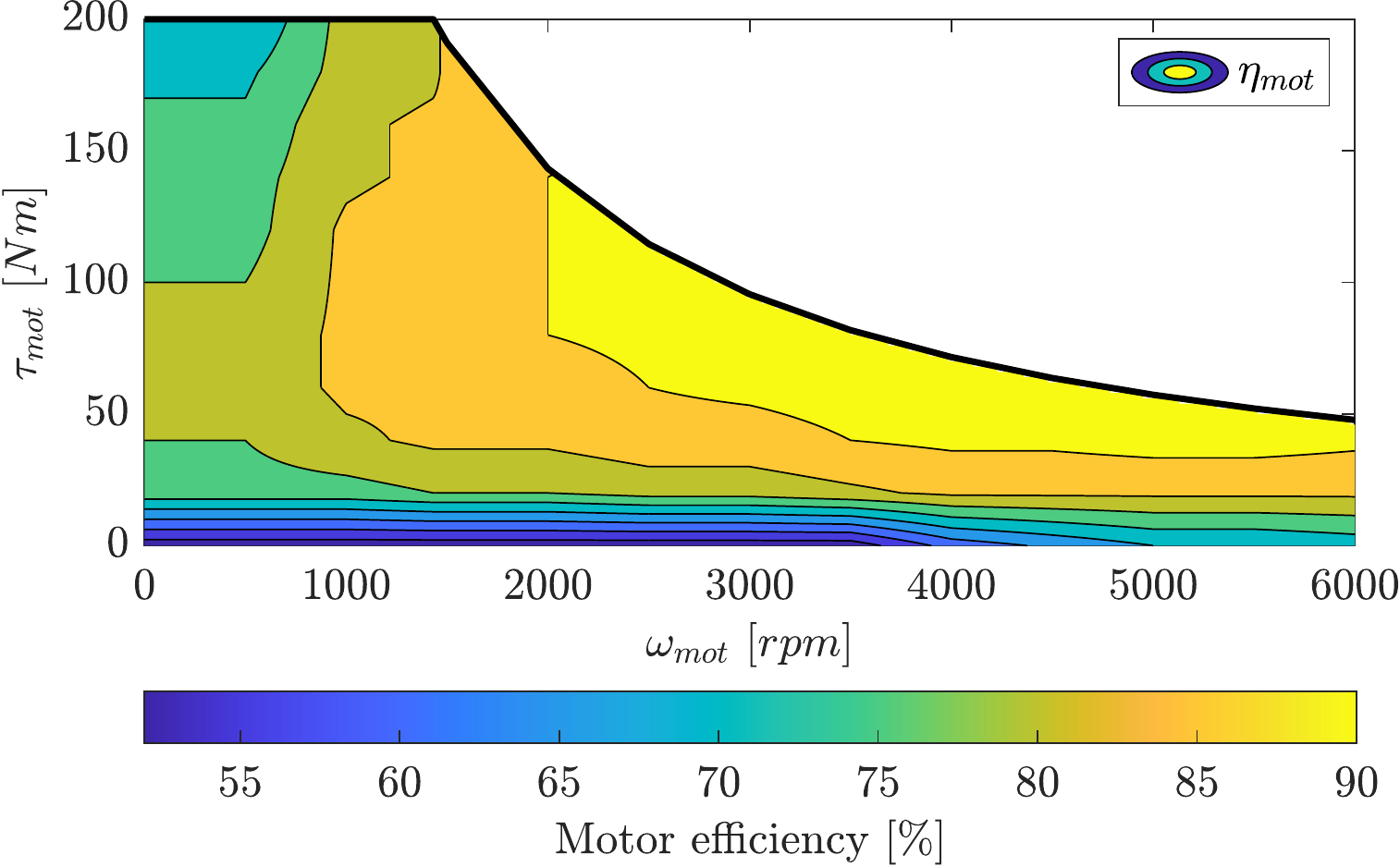}}
	\caption{Motor efficiency maps for EV and HEV.}
	\label{fig:motor_map}	
\end{figure}

Electric motors are energy conversion devices working either in motoring (i.e., \change{torque is provided to the wheels}) or in generating (i.e., \change{torque is received from the wheels}) mode. \changeb{For the purpose of this work, the actuator is modeled using static efficiency maps $\eta_{mot}(\tau_{mot},\omega_{mot})$, function of the motor torque ($\tau_{mot}$) and speed ($\omega_{mot}$). The static maps used in this work, for the EV and HEV case studies, are shown in Fig. \ref{fig:motor_map}.} The motor torque is in turn obtained as
\begin{equation}
	\tau_{mot}=\textrm{min}\left(\left|\frac{P_{mot}(t)}{\omega_{mot}}\right|,\tau_{max,mot}\right),
\end{equation}
where  $\tau_{max,mot}$ is the maximum torque that the motor can deliver, and the motor power, $P_{mot}$, is computed as
\begin{equation}
	P_{mot}=P_{batt}\cdot \eta^{\mathrm{sign}(I_{batt})}_{mot},
\end{equation}
\change{with the $\mathrm{sign}$ function defined as:
\begin{equation}
\mathrm{sign}(I_{batt})=\begin{array}{c}
\begin{cases}
-1\ &\text{if}\ I_{batt} < 0,\\
0\ &\text{if}\ I_{batt} = 0,\\
1\ &\text{if}\ I_{batt} > 0.
\end{cases}\end{array}
\end{equation}}
Following the same reasoning of the battery, a thermal model for the motor is introduced. The thermal model, as well as its parameters, are borrowed from \cite{rajput_electric_motor}. This reference provides data for Interior Permanent Magnet Synchronous Machines (IPMSMs), commonly used devices \changeb{in both EVs and HEVs} \cite{zhu_motors}. \changeb{The model describes the temperature evolution of both copper windings and stator iron.} The heat generation within the rotor is considered negligible and the motor is assumed to directly exchange heat with the environment (no coolant is considered).
The model is characterized by two heat capacities $\mathrm{C}_{mot,copper}$ and $\mathrm{C}_{mot,iron}$, as well as by two thermal transfer coefficients $\mathrm{h}_{mot,copper}$ and $\mathrm{h}_{mot,iron}$, for copper and iron, respectively. These coefficients are combined as follows
\begin{equation}
\begin{split}
&\mathrm{C}_{mot}=\alpha \cdot \mathrm{C}_{mot,copper}+\beta\cdot \mathrm{C}_{mot,iron}, \\
&\mathrm{h}_{out,mot}=\alpha\cdot \mathrm{h}_{mot,copper}+\beta\cdot \mathrm{h}_{mot,iron},
\end{split}
\end{equation}
where $\alpha$ and $\beta$ represent the copper and iron mass fractions in the device, set to $0.15$ and $0.85$ \cite{goss2013comparison}, respectively. \changeb{The motor losses account for Joule effect (in the copper phase), iron hysteresis, and friction}
\begin{equation}
\resizebox{1\columnwidth}{!}
	{$\begin{split}
	\label{eq:mot_losses}
&	P_{mot,copper}(t)=R_s(T_{mot}(t))\cdot \left(I_d^2(t)+I_q^2(t)\right), \\
&   P_{mot,iron}(t)=k_h\cdot \omega_{mot}\cdot\left[\left(L_d\cdot I_d+\Lambda_{pm}\right)^2+\left(L_q\cdot I_q\right)^2\right], \\
& P_{mot,fric}(t)=k_f\cdot \omega^2_{mot},
\end{split}$}
\end{equation}
where $I_d$ and $I_q$ are the $d$ and $q$ axes currents, $L_d$ and $L_q$ are the $d$ and $q$ axes inductances, $k_h$ and $k_f$ are experimentally obtained parameters used to compute the \changeb{motor iron and friction losses, and $\Lambda_{pm}$ is defined as $\sqrt{3/2}\cdot \lambda_{pm}$ (with $\lambda_{pm}$ the permanent magnet flux linkage).} $R_s$ is the stator resistance, function of the motor temperature $T_{mot}$ 
\begin{equation}
	R_s=R_{s,0}\cdot \left[1+\xi \cdot \left(T_{mot}-T_0\right) \right],
\end{equation}
\changeb{where $R_{s,0}$ is the stator resistance at $T_0$ and $\xi$ is an identified parameter modeling the temperature dependence.} The computation of $I_d$ and $I_q$ would require the simulation of the low level electrical dynamics and controls of the motor. Since this is out of the scope of the work, a simplified procedure for the computation of these currents is exploited. We assume the motor to be controlled \changeb{by} a maximum torque per ampere (MTPA) algorithm, which, as described in \cite{mtpa}, provides the $I_{d,ref}$ and $I_{q,ref}$ \changeb{reference currents} to the low level controller. Then, assuming the reference currents to be perfectly tracked, \changeb{the following holds}
\begin{equation}
I_{d}=I_{d,ref},\quad I_{q}=I_{q,ref}.
\end{equation}
Overall, the electric motor thermal model reads
\begin{equation}
\begin{split}
	&\mathrm{C}_{mot}\cdot \dot{T}_{mot}=\mathrm{h}_{out,mot}\cdot \left(T_0-T_{mot}\right)+P_{mot,fric}+\\&+P_{mot,copper}+P_{mot,iron}.
\end{split}
\label{eq:motor_thermal}
\end{equation}

Similarly to the battery pack, \changeb{the electric motor is a closed and incompressible system and the exergy balance is written as}
\begin{equation}
	\dot{X}_{mot}(t)=\dot{X}_{heat,mot}(t)+\dot{W}_{mot}(t)+\dot{X}_{dest,mot}(t),
	\label{eq:motor_exergy}
\end{equation}
where $\dot{X}_{mot}$ is the rate of exergy change in the motor and $\dot{X}_{heat,mot}$ is the exergy transfer rate due to heat, computed as
\begin{equation}
\dot{X}_{heat,mot}=\left(1-\dfrac{T_0}{T_{mot}}\right)\cdot \dot{Q}_{mot}(t),
\label{eq:exergy_heat_trans}
\end{equation}
with $\dot{Q}_{mot}=\mathrm{h}_{out,mot}\cdot(T_0-T_{mot})$ the heat exchange between the motor and the environment. $\dot{W}_{mot}$ is the work \change{rate} related to the motor, which is equal to zero. This is reasonable because both \change{the works at the input and output} of the motor are already taken into account in other components of the powertrain: the work in input to the motor is the battery power $P_{batt}$, and the work in output from the motor is the traction power $P_{trac}$ of the longitudinal dynamics model (as defined in Section \ref{section:long_dynamics}).

\changeb{The term} $\dot{X}_{dest,mot}=-T_0\cdot \dot{S}_{gen,mot}(t)$ is the rate of exergy destruction within the motor, and $\dot{S}_{gen,mot}$ is the related entropy generation computed according to the procedure shown for the battery \change{(see Equation (\ref{eq:entropy_rate_battery}))}. Assuming the \change{electric motor} to be incompressible and with constant heat capacity, the following balance is obtained
\begin{equation}
\begin{split}
\dot{S}_{mot}(t)=&\ \dot{S}_{gen,mot}+\dot{S}_{in,mot}(t)-\dot{S}_{out,mot}(t),
\label{eq:entropybalance_motor}
\end{split}
\end{equation}
where $\dot{S}_{mot}(t)$ is the motor entropy rate, and \change{$\dot{S}_{in,mot}$ and $\dot{S}_{out,mot}$} are the entropy transfer into or \changeb{out of} the motor, respectively. The \changeb{last term on the right-hand-side of Equation (\ref{eq:entropybalance_motor})} is computed as
\begin{equation}
	-\dot{S}_{out,mot}=\dfrac{\dot{Q}_{mot}}{T_{mot}}=\mathrm{h}_{out,mot}\cdot \dfrac{T_0-T_{mot}}{T_{mot}}.
	\label{eq:heat_transf_motor}
\end{equation}
$\dot{S}_{in,mot}$ is equal to zero because $T_{mot}>T_0,\ \forall t$. \changeb{Moreover}, following the same procedure showed in \changeb{Equations} (\ref{eq:battery_entropy_1}) and (\ref{eq:battery_entropy_2}), the motor entropy rate reads as
\begin{equation}
	\dot{S}_{mot}=\dfrac{\mathrm{C}_{mot}\cdot {\dot{T}_{mot}}}{T_{mot}},
	\label{eq:entropy_motor_1}
\end{equation}
\changeb{and the motor entropy generation is obtained combining Equations (\ref{eq:entropybalance_motor}), (\ref{eq:heat_transf_motor}), and (\ref{eq:entropy_motor_1})}
\begin{equation}
	\dot{S}_{gen,mot}=\dfrac{P_{mot,copper}+P_{mot,iron}+P_{mot,fric}}{T_{mot}}
	\label{eq:motor_entropy_gen}.
\end{equation}

Finally, the motor exergy balance is \changeb{obtained} substituting Equations (\ref{eq:exergy_heat_trans}) and (\ref{eq:motor_entropy_gen}) into Equation (\ref{eq:motor_exergy}) \changeb{as}
\begin{equation}
\begin{split}
\dot{X}_{mot} &= \left(1-\frac{T_0}{T_{mot}}\right)\cdot\dot{Q}_{mot}-\\
                             &\hspace{1em}-\frac{T_0\cdot(P_{mot,copper}+P_{mot,iron}+P_{mot,fric})}{T_{mot}}.
\end{split}
\label{eq:motor_fullex}
\end{equation}

\subsubsection{Internal Combustion Engine}
\label{subsubsection:engine_exergy}
 \begin{figure}[!t]
	\includegraphics[width=\columnwidth]{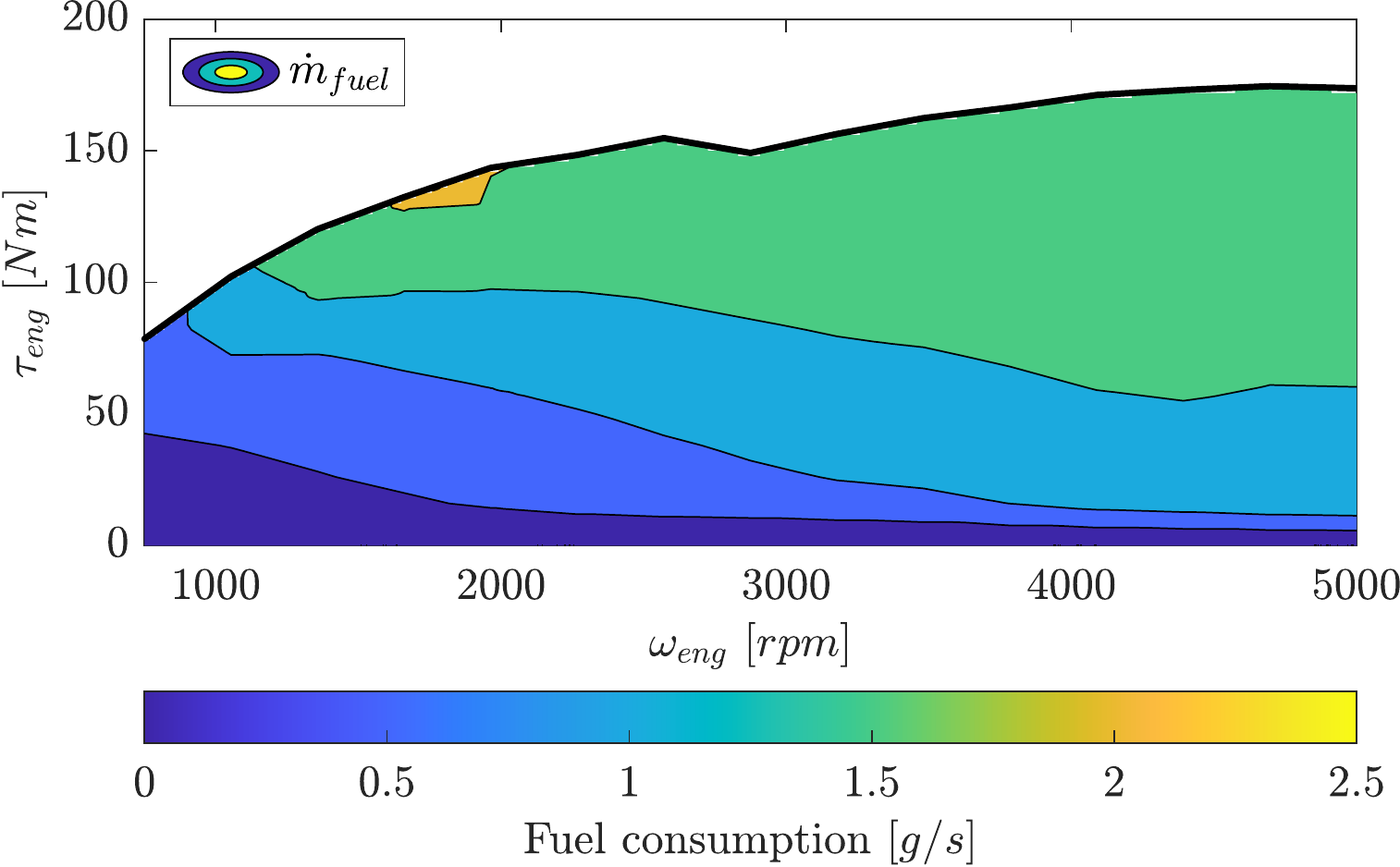}
	\caption{\changeb{ICE fuel consumption map used in this work.}}
	\label{fig:ice_eff_map}	
\end{figure}
\change{In this work, an inline $4$-cylinder gasoline engine is considered \changeb{for use in the HEV architecture}.} \change{Starting from \cite{razmara_exergy_engine}, the steady-state exergy balance for the ICE is written as follows}
\begin{equation}
\label{eq:eng_exergy_1}
\begin{split}
&\dot{X}_{fuel,eng}(t)+\dot{X}_{intk,eng}(t)=-\left[\dot{X}_{work,eng}(t)+\right. \\
&\left.\dot{X}_{exh,eng}(t)+\dot{X}_{heat,eng}(t)+\dot{X}_{fric,eng}(t)+\right. \\
&\left.+\dot{X}_{comb,eng}(t)\right],
\end{split}
\end{equation}
where $\dot{X}_{fuel,eng}$ is the \change{fuel availability}, computed according to \cite{sayin_exergy_engine}
\begin{equation}
\resizebox{1\columnwidth}{!}
	{$\begin{split}
\dot{X}_{fuel,eng}&=X_{spec,fuel}\cdot \dot{m}_{fuel}(t)=\\
&=\left(1.04224+0.011925\cdot \dfrac{x}{y}-\dfrac{0.0042}{x}\right)\cdot LHV \cdot \dot{m}_{fuel}(t).
\end{split}$}
\label{eq:fuel_exergy}
\end{equation}
with $X_{spec,fuel}$ the specific fuel chemical exergy, $x$ and $y$ known from the fuel chemical formula $C_{x}H_{y}$ (e.g., for gasoline $x=8$, $y=18$), and $LHV$ the fuel lower heating value. \change{The fuel consumption $\dot{m}_{fuel}$ is retrieved from the map in Fig. \ref{fig:ice_eff_map}, function of $\omega_{eng}$ -- the engine rotational speed -- and $\tau_{eng}$ -- the engine torque --.} When the tank is completely filled, the maximum fuel availability is computed as
\begin{equation}
	X_{fuel,max}=\mathcal{V}_{tank}\cdot \rho_{fuel}\cdot X_{spec,fuel},
	\label{eq:max_avail_fuel}
\end{equation}
where $\mathcal{V}_{tank}$ is the tank volume and $\rho_{fuel}$ is the fuel density. \changeb{$\dot{X}_{intk,eng}$ is the intake exergy flow, related to the air entering the engine for combustion, and, according to Equation (\ref{eq:overal_ex_theory}), takes the following form}
\begin{equation}
\dot{X}_{intk,eng} = \sum_{j\in\mathcal{K}}\dot{n}_{intk,j}(t)\cdot \psi_{j}(t).
\end{equation}
where $\dot{n}_{intk,j}$ is the molar flow rate of a species $j$ in the intake manifold and $\psi_{j}$ is the exergy flux. This term accounts for less than  1\% of the fuel availability and can be neglected \cite{sayin2016energy}: $\dot{X}_{intk,eng} = 0$. 

\change{The term $\dot{X}_{exh,eng}$ models the exergy exchanged with the environment through the exhaust gas. According to Section \ref{section:theory}, \change{this term is function of the species $k$ composing the exhaust gas, namely, $N_2$, $O_2$, $H_2O$, and $CO_2$\footnote{Other species, i.e., $CO$, $NO_x$, and argon, are present in small concentrations (the volume fraction is $<0.01$) and can be neglected \cite{guzzella_book}.}, and is defined as}
\begin{equation}
\begin{split}
\dot{X}_{exh,eng}&=-\sum_{k\in\mathcal{K}}\dot{n}_{exh,k}(t)\cdot \psi_{k}(t)=\\
&=-\sum_{k\in\mathcal{K}}\dot{n}_{exh,k}\cdot \left(\psi_{ch,k}(t)+\psi_{ph,k}(t)\right),
\end{split}
\label{eq:ex_exhaust}
\end{equation}
where $\dot{n}_{exh,k}$ is the molar flow rate of a species $k$ in the exhaust manifold and $\psi_{ch,k}$ and $\psi_{ph,k}$ are the specific chemical and physical exergies \changeb{defined in Section \ref{section:theory}}. To describe the combustion process and compute Equation (\ref{eq:ex_exhaust}), we \changeb{exploit} the mean-value approach, in which the combustion process is not analyzed cycle by cycle (i.e., in the crank-angle domain), but averaged, over time, for the \changeb{four cylinders}. This is a reasonable approach since the goal \changeb{of this work} is the exergetic characterization of the ICE and not the optimization of the combustion process variables, such as spark, injection, and valve timings. Assuming the engine \changeb{is working} at stoichiometric conditions, all the oxygen is burnt during combustion according to the following reaction \cite{heywood_book}
\begin{equation}
\begin{split}
&C_8H_{18}+z_{comb} \left(O_2+3.76N_2\right)\rightarrow\\
&a_{comb}CO_2+b_{comb}H_2O+c_{comb}N_2,
\end{split}
\label{eq:comb_reaction}
\end{equation}
where $a_{comb}=x$, $b_{comb}=y/2$, $z_{comb}=a_{comb}+b_{comb}/2$ and $c_{comb}=3.76z_{comb}$. The stoichiometric assumption is realistic for a spark-ignition engine and ensures the optimal operation of the three-way catalyst \cite{heywood_book}. Starting from the fuel mass flow rate, $\dot{m}_{fuel}$, known for a given engine operating point, the air mass flow rate \changeb{is}
\begin{equation}
	\dot{m}_{air}(t)=\dot{m}_{fuel}\cdot AFR_{stoich},
\end{equation}
with $AFR_{stoich}$ the stoichiometric air-fuel ratio. In accordance with \cite{sayin_exergy_engine}, the exhaust gas mass flow rate is computed as $\dot{m}_{exh}(t)=\dot{m}_{fuel}+\dot{m}_{air}$ and the corresponding molar flow rate reads as
\begin{equation}
\dot{n}_{exh}(t)=\dfrac{\dot{m}_{exh}}{\sum_{k\in\mathcal{K}}f^\star_{exh,k}\cdot M_{k}},
\label{eq:nexh}
\end{equation}
where $M_{k}$ is the molar mass for the $k$-th species in the exhaust manifold. Starting from Equation (\ref{eq:nexh}), the contribution of the different exhaust gas species to $\dot{n}_{exh}$ is given by
\begin{equation}
	\dot{n}_{exh,k}(t)=\dot{n}_{exh}\cdot f^\star_{exh,k},
\end{equation}
where $f^\star_{exh,k}$ is the molar fraction of the species $k$ at the restricted state 
\begin{equation}
\begin{split}
	&f^\star_{exh,CO_2}=\dfrac{a_{comb}}{n_{exh,tot}},\ f^\star_{exh,H_2O}=\dfrac{b_{comb}}{n_{exh,tot}},\\ &f^\star_{exh,N_2}=\dfrac{c_{comb}}{n_{exh,tot}},\ f^\star_{exh,O_2}=0,
\end{split}
\end{equation}
with $n_{exh,tot}=a_{comb}+b_{comb}+c_{comb}$ the number of moles of the products, from the right hand side of Equation (\ref{eq:comb_reaction}). 
The chemical exergy is obtained \changeb{similarly to} Equation (\ref{eq:chem_exergy})
\begin{equation}
\psi_{ch,k}=\mathcal{R}_{gas}\cdot T_0\cdot \log \left(\dfrac{f^\star_{exh,k}}{f_{k,0}}\right).
\label{eq:eng_exh_chem}
\end{equation}
Recalling Equation (\ref{eq:phys_ex_general}), the physical exergy \changeb{is}
\begin{equation}
\psi_{ph,k} = h_k - h_k^\star-T_0\cdot(s_k-s_k^\star),
\label{eq:eng_exh_phys}
\end{equation} 
where $h_k$ and $h^\star_{k}$ are the specific enthalpies of a species $k$ in the exhaust manifold and at the restricted state, respectively. Similarly, $s_k$ and $s^\star_{k}$ are the specific entropies in the exhaust manifold and at the restricted state. To compute the thermodynamic properties of the gaseous species (enthalpy and entropy), the experimentally fitted NASA polynomials \cite{burcat_polynomials}, function of the exhaust gas mixture temperature, are used.}
\begin{figure*}
	\begin{equation}
	\resizebox{1\textwidth}{!}
	{$\begin{split}
		\boldsymbol{X^{EV}_{veh}} &= \int_{0}^{t_f} \boldsymbol{\dot{X}^{EV}_{veh}}\ dt =\int_{0}^{t_f} \left(\textcolor{Cerulean}{\dot{X}_{long}}+\textcolor{BurntOrange}{\dot{X}_{batt}}+\textcolor{Green}{\dot{X}_{mot}}\right)\ dt=\\
		&=\int_{0}^{t_f}\left(\textcolor{Cerulean}{P_{trac}^{EV}-P_{brake}-P_{roll}-P_{aero}}\right)+\left(\textcolor{BurntOrange}{-P_{batt}+\dot{X}_{dest,batt}+ \dot{X}_{heat,batt}}\right)+2\cdot\left(\textcolor{Green}{\dot{X}_{heat,mot}+\dot{X}_{dest,mot}}\right)\ dt=\\
		&=X_{batt}(0)+\left(\textcolor{Cerulean}{E_{trac}^{EV}-E_{brake}-E_{roll}-E_{aero}}\right)+\left(\textcolor{BurntOrange}{-E_{batt}+X_{dest,batt}+X_{heat,batt}}\right)+2\cdot\left(\textcolor{Green}{X_{heat,mot}+X_{dest,mot}}\right).
		\end{split}$}
	\label{eq:exergy_EV}
	\end{equation}
\end{figure*}
\begin{figure*}
	\begin{equation}
	\resizebox{0.95\textwidth}{!}
	{$\begin{split}
	\boldsymbol{X^{HEV}_{veh}} &= X_{fuel,max}+\int_{0}^{t_f}\boldsymbol{\dot{X}^{HEV}_{veh}}dt=X_{fuel,max}+\int_{0}^{t_f}\left(\textcolor{Cerulean}{\dot{X}_{long}}+\textcolor{BurntOrange}{\dot{X}_{batt}}+\textcolor{Green}{\dot{X}_{mot}}-\textcolor{OrangeRed}{\dot{X}_{fuel,eng}}\right)dt=\\
	&= X_{fuel,max}+X_{batt}(0)+\int_{0}^{t_f}\left(\textcolor{Cerulean}{P_{trac}^{HEV}-P_{brake}-P_{roll}-P_{aero}}\right)+\left(\textcolor{BurntOrange}{-P_{batt}+\dot{X}_{dest,batt}+\dot{X}_{heat,batt}}\right)+\\
	&\hspace{1em}+\left(\textcolor{Green}{\dot{X}_{heat,mot}+\dot{X}_{dest,mot}}\right)+\left(\textcolor{OrangeRed}{\dot{X}_{comb,eng}+\dot{X}_{heat,eng}+\dot{X}_{work,eng}+\dot{X}_{fric,eng}+\dot{X}_{exh,eng}}\right)dt=\\
	&=X_{fuel,max}+X_{batt}(0)+\left(\textcolor{Cerulean}{E_{trac}^{HEV}-E_{brake}-E_{roll}-E_{aero}}\right)+\left(\textcolor{BurntOrange}{-E_{batt}+X_{dest,batt}+X_{heat,batt}}\right)+\\
	&\hspace{1em}+\left(\textcolor{Green}{X_{heat,mot}+X_{dest,mot}}\right)+\left(\textcolor{OrangeRed}{X_{comb,eng}+X_{heat,eng}+X_{work,eng}+X_{fric,eng}+X_{exh,eng}}\right).
	\end{split}$}
	\label{eq:exergy_HEV}
	\end{equation}
\end{figure*}

The exergy transfer related to the \change{mechanical work generation} ($\dot{X}_{work,eng}$) is obtained as follows
\begin{equation}
\dot{X}_{work,eng}=-P_{eng}(t)=-\tau_{eng}(t)\cdot \omega_{eng}(t).
\label{eq:eng_work}
\end{equation}
In accordance with Equation(\ref{eq:generic_exheat}), the availability change related to heat transfer \change{towards the} cylinder walls is modeled as
\begin{equation}
\dot{X}_{heat,eng}=\left(1-\dfrac{T_0}{T_{eng}}\right)\cdot \dot{Q}_{eng}(t),
\label{eq:eng_heat}
\end{equation}
where $\dot{Q}_{eng}$ is the thermal exchange between the in-cylinder mixture, at temperature $T_{eng}$, and the cylinder walls. Relying on the time-averaged Taylor\&Toong correlation \cite{taylor1985internal}, the heat transfer $\dot{Q}_{eng}$ is computed as follows
\begin{equation}
\resizebox{1\columnwidth}{!}{$
\dot{Q}_{eng}=a\frac{k_g}{\mu_g^b}(\dot{m}_{fuel}+\dot{m}_{air})^b B^{b-1}\left(\frac{\pi B^2}{4}\right)^{1-b}(T_{eng}-T_{c}),$}
\label{eq:ice_thermal}
\end{equation}
where $k_g$ is the mixture conductivity, $\mu_g$ the mixture viscosity, $B$ the cylinder bore, $T_{c}$ the coolant temperature, and $a$ and $b$ empirical, dimensionless, coefficients function of the engine characteristics: $a$ is tuned in order to reach a contribution of $\dot{X}_{heat,eng}$ to the balance in Equation (\ref{eq:eng_exergy_1}) of around $10\%$ (in line with \cite{rakopoulos_exergy_engine}). $T_{eng}$, $k_g$, and $\mu_g$ are function of the air-fuel ratio which, in this work, is constant and equal to the stoichiometric value $AFR_{stoich}$. Therefore, $T_{eng}$, $k_g$, and $\mu_g$ are also constant and determined relying on data available in \cite{heywood_book} for spark-ignition engines. 

\changeb{Finally, the combustion irreversibility term, the principal source of loss in the ICE \cite{caton_engine,razmara_exergy_engine}, is obtained inverting Equation (\ref{eq:eng_exergy_1})}
\begin{equation}
\begin{split}
\dot{X}_{comb,eng}=&-\dot{X}_{fuel,eng}-\dot{X}_{work,eng}-\\
&-\dot{X}_{exh,eng}-\dot{X}_{heat,eng}-\\
&-\dot{X}_{fric,eng}.
\end{split}
\label{eq:eng_irrev}
\end{equation}
with $\dot{X}_{fric,eng}$ the friction exergy loss computed \change{numerically}. \changeb{Through Equation (\ref{eq:eng_irrev}), one avoids to compute the complex governing combustion reactions. Finally,} substituting Equations (\ref{eq:ex_exhaust}),  (\ref{eq:eng_work}), (\ref{eq:eng_heat}), (\ref{eq:eng_irrev}), into Equation (\ref{eq:eng_exergy_1}), the engine exergy balance can be obtained.

\subsection{Vehicle Exergy Balance}\label{subsec:overall_bal}
Combining the \change{exergy rate expressions} derived for the \change{electrochemical storage device} (Section \ref{section:batt_electric_vehicle}), the electric motor (Section \ref{section:EM_electric_vehicle}), the ICE (Section \ref{subsubsection:engine_exergy}), and the vehicle longitudinal dynamics (Section \ref{section:long_dynamics}), the overall exergy balance for EVs and HEVs is derived \changeb{and reported in Equations (\ref{eq:exergy_EV}) and (\ref{eq:exergy_HEV}), respectively}. The formulation of the longitudinal dynamics term, $\dot{X}_{long}$, is the same for both architectures, with the traction power at the wheels computed \changeb{either as
\begin{equation}
P^{EV}_{trac}= P_{mot}\cdot\eta_{diff}^{\mathrm{sign}(I_{batt})} ,
\end{equation}
or
\begin{equation}
P^{HEV}_{trac}= (P_{mot}+P_{eng})\cdot\eta_{diff}^{\mathrm{sign}(I_{batt})},
\end{equation}
whether the vehicle is an EV or HEV. In particular, for the EV case, the balance in Equation (\ref{eq:exergy_EV}) is given by summing} to $\dot{X}_{long}$ the battery and electric motor exergy rates. \changeb{In this context, two identical electric motors are used and the corresponding exergy term is multiplied by a factor of two. A similar} procedure is followed for the HEV, where $\dot{X}_{veh}^{HEV}$ is obtained \changeb{including} also the ICE (Equation (\ref{eq:exergy_HEV})). Therefore, the rates are integrated over the length of the driving cycle, $t_f$, and a quantification of the powertrain exergy is obtained. In Equations (\ref{eq:exergy_EV}) and (\ref{eq:exergy_HEV}), the term $X_{batt}(0)$ is the exergy stored in the battery at the beginning of a driving cycle, function of the initial $SoC$ \changeb{and of the battery pack nominal energy $E_{nom}$}
\begin{equation}
X_{batt}(0) = SoC(0)\cdot E_{nom}.
\end{equation}
In Equation (\ref{eq:exergy_HEV}), the term $X_{fuel,max}$ represents the exergy initially stored in the fuel tank (defined in Equation (\ref{eq:max_avail_fuel})). \changeb{To facilitate the reading, Equations (\ref{eq:exergy_EV}) and (\ref{eq:exergy_HEV}) are color-coded. Light-blue, orange, green, and magenta are used for the exergy terms associated with the longitudinal dynamics, battery, electric motor, and ICE, respectively.}

\change{In the balances $\dot{X}_{veh}^{EV}$ and  $\dot{X}_{veh}^{HEV}$, $\dot{X}_{batt}$} represents a storage feature describing the way the stored exergy is exchanged with the environment ($\dot{X}_{heat,batt}$), transformed into useful work ($P_{batt}$), or destroyed ($\dot{X}_{dest,batt}$). On the other hand, \change{$\dot{X}_{long}$ and $\dot{X}_{mot}$ describe how} exergy is transferred to the wheels ($P_{trac}$), destroyed ($P_{aero}$, $P_{roll}$, $P_{brake}$, or $\dot{X}_{dest,mot}$), or exchanged with the environment ($\dot{X}_{heat,mot}$). \changeb{In the case of the HEV,} $\dot{X}_{fuel,eng}$ is the exergy stored in the fuel, which is then transformed into useful work ($\dot{X}_{work,eng}$), destroyed (e.g., through $\dot{X}_{comb,eng}$), and exchanged with the environment (e.g., through $\dot{X}_{heat,eng}$). 

As soon as the vehicle starts moving, the exergy \change{flows} from the energy sources to the wheels and is lost due to transfer and destruction phenomena. The vehicle exergy can never increase \changeb{over} a driving cycle, it can only be destroyed, lost to the environment, or used to propel the vehicle. \changeb{To exploit this concept, the following normalized quantities are defined}
\begin{equation}
\begin{split}
	X^{EV}_{rel}&=\dfrac{X^{EV}_{veh}}{X_{batt,max}},\\
	X^{HEV}_{rel}&=\dfrac{X^{HEV}_{veh}}{X_{batt,max}+X_{fuel,max}},
\end{split}
\label{eq:ex_normalization}
\end{equation}
\changeb{where $X_{batt,max}=E_{nom}$ and, as shown in Equations (\ref{eq:exergy_EV}) and (\ref{eq:exergy_HEV}), $X^{EV}_{veh}$ and $X^{HEV}_{veh}$ are the EV and HEV exergy states, respectively. For the EV case, $X^{EV}_{rel} = 1$ corresponds to the maximum available work, i.e., to the fully charged battery. On the other hand, in the HEV the availability is maximized when the battery pack is fully charged and the tank is filled to \changeb{its} maximum capacity: this condition corresponds to $X^{HEV}_{rel} = 1$. }

\changeb{The net amount of exergy lost in the powertrain due to the conversion of the electrical (from the battery) and mechanical (from the ICE) power into traction power is referred to as $E_{loss,pwt}$ and takes the following form
\begin{equation}
\begin{split}
	E^{EV}_{loss,pwt}&=E_{trac}^{EV}-E_{batt},\\ E^{HEV}_{loss,pwt}&=E_{trac}^{HEV}-E_{batt}+X_{work,eng},
\end{split}
\label{eq:pwt_losses}
\end{equation}
for the EV and HEV case, respectively. In Equation (\ref{eq:pwt_losses}), $E_{trac}^{EV}$, $E_{trac}^{HEV}$, and $E_{batt}$ are obtained integrating the corresponding quantities $P_{trac}^{EV}$, $P_{trac}^{HEV}$, and $P_{batt}$ over the duration of the driving cycle.}

\section{Case Studies}
\label{section:case_study}
\begin{figure}[!t]
	\centering
	\subfloat[\label{fig:ev_architecture}Electric Vehicle.]{\includegraphics[width=0.9\columnwidth]{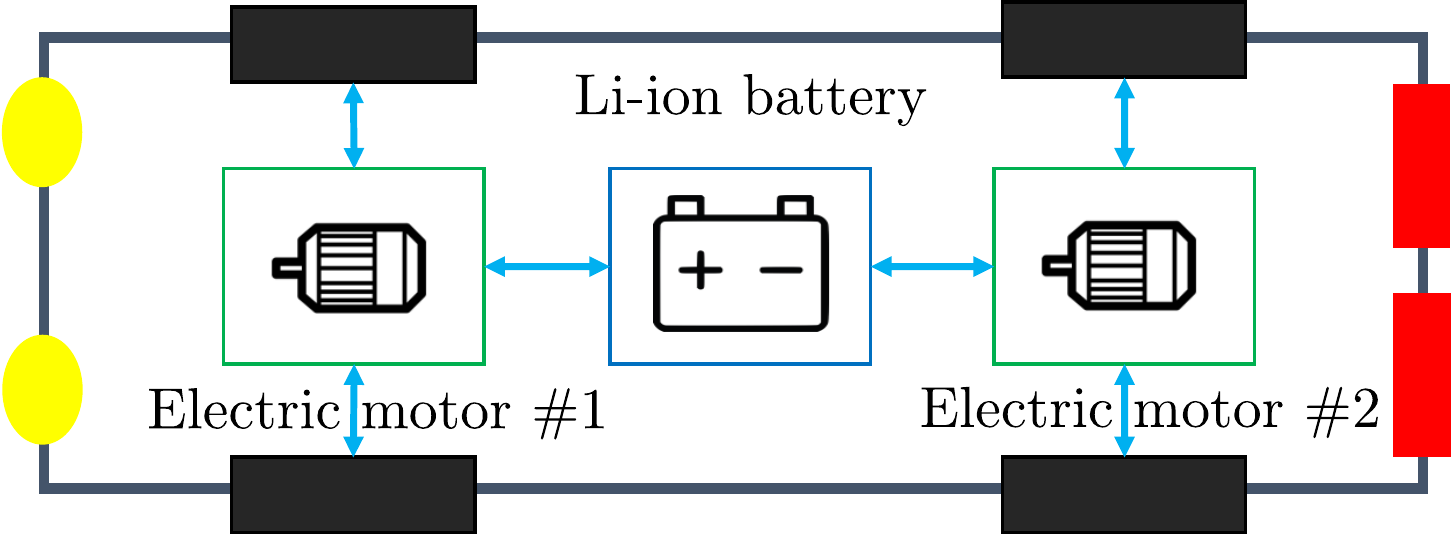}}\\
	\subfloat[\label{fig:hev_architecture}Hybrid Electric Vehicle.]{\includegraphics[width=.9\columnwidth]{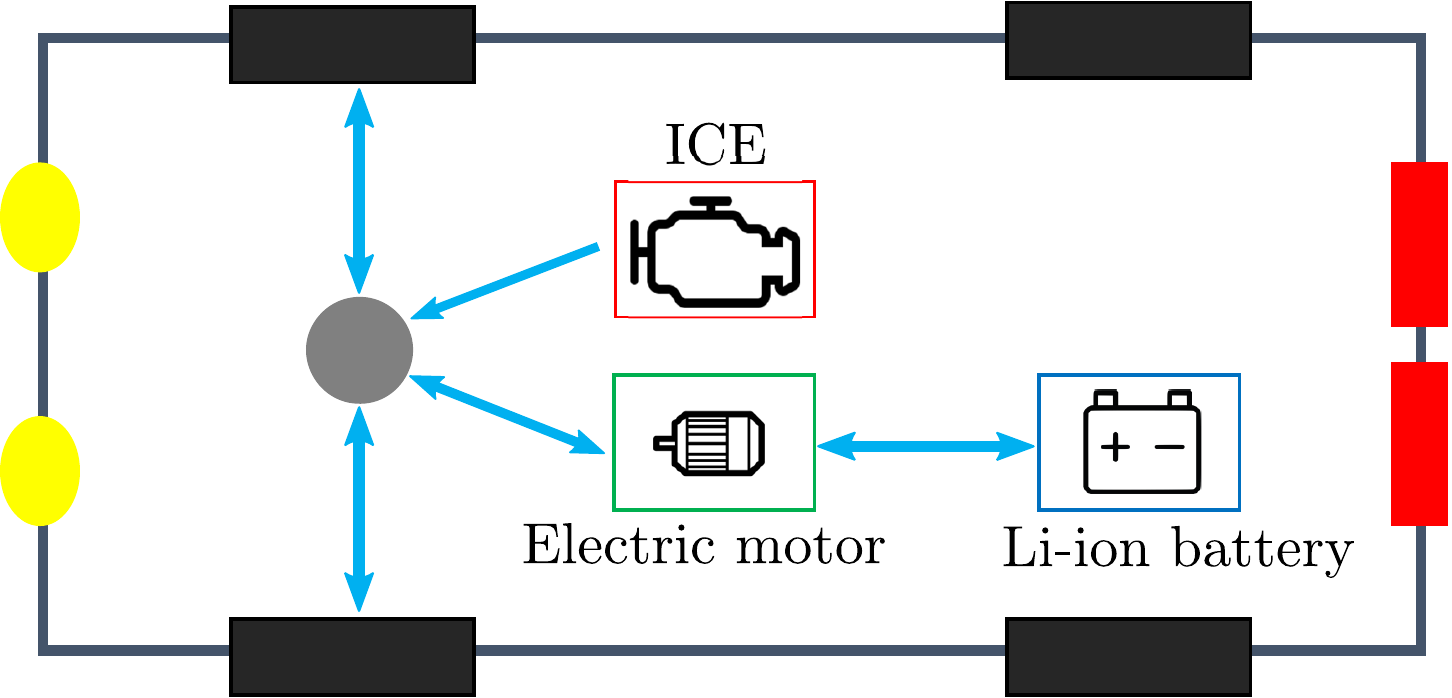}}
	\caption{Schematic representation of the powertrain architectures. }
	\label{fig:vehicle_architecture}
\end{figure}

\changeb{The proposed framework is tested on an EV and a parallel HEV characterized by the parameters listed in Table \ref{tab:vehicle_parameters_table}. The EV is equipped with two electric motors (with efficiency map in Fig. \ref{fig:motor_map}a) and a battery pack with nominal energy $E_{nom} = 90kWh$. The HEV has one electric motor (with efficiency map in Fig. \ref{fig:motor_map}b), a battery pack with nominal energy $E_{nom} = 1kWh$, and a spark-ignition ICE characterized by the fuel consumption map in Fig. \ref{fig:ice_eff_map}. A schematic representation of the two architectures is provided in Fig. \ref{fig:vehicle_architecture}.}

The simulators are borrowed from the MathWorks Powertrain Blockset toolbox \cite{powertrainblockset}, in which forward models for both EVs and HEVs are provided. The driving cycle \changeb{is} the driver's desired speed, which is followed relying on the battery (for the EV) or on a combination of battery and ICE for the HEV case study (a tracking error lower than $1km/h$ is ensured). \changeb{The \textsc{Matlab} model is enhanced with the battery, electric motor, and ICE thermal models -- described in Equations (\ref{eq:upscaledall}), (\ref{eq:motor_thermal}), and (\ref{eq:ice_thermal}) -- and the corresponding exergy balances -- formalized in Equations (\ref{eq:battery_fullex}), (\ref{eq:motor_fullex}), and (\ref{eq:eng_exergy_1}) --.}

\change{Simulations are carried out considering the World harmonized Light-duty vehicles Test Procedure (WLTP), featuring a mix of urban and highway driving conditions \cite{update2013world}}, and the \changeb{reference state} defined in Section \ref{section:ref_env}. 
\begin{figure}[!tb]
	\centering
	\subfloat[Speed and C-rate profiles for the EV.]{\includegraphics[width=0.92\columnwidth]{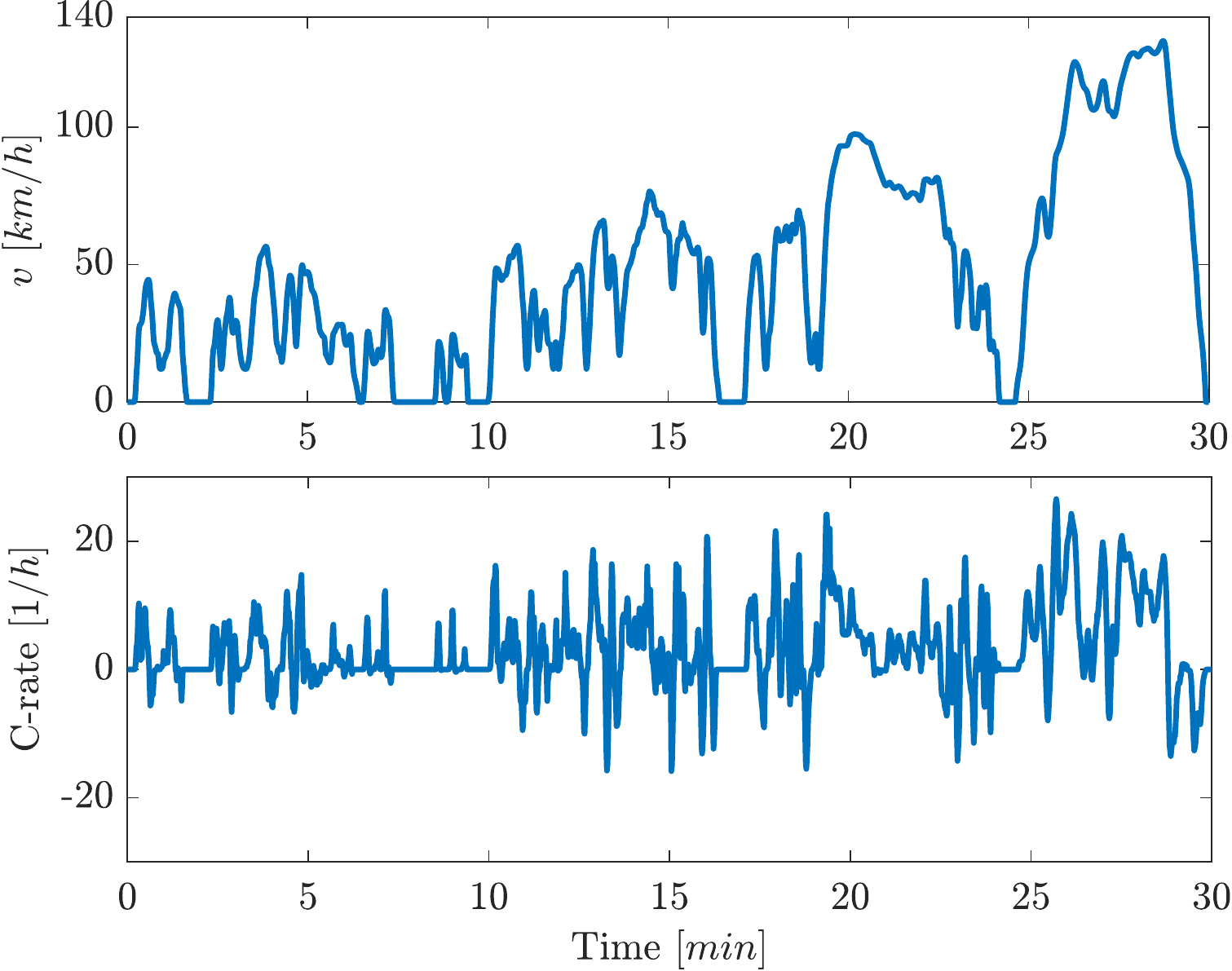}}
	\hfill
	\subfloat[Speed, C-rate, and electric motor/engine torques for the HEV. \label{fig:wltp_cycle_hev}]{\includegraphics[width=0.92\columnwidth]{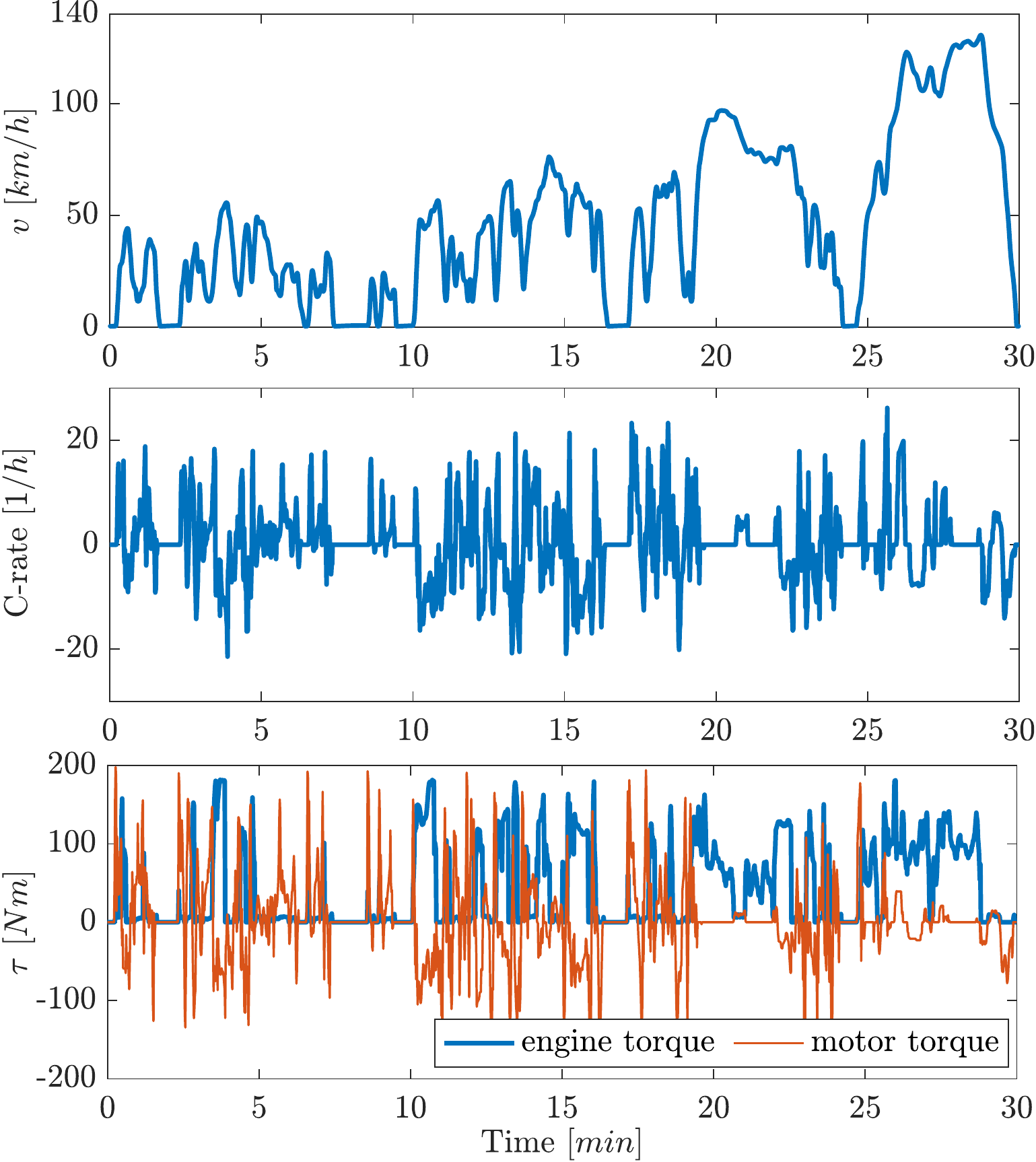}}
	\caption{Simulation results for the EV and HEV \change{along the WLTP driving cycle.}\label{fig:wltp_cycle}}
\end{figure}
\begin{figure*}[!htb]
	\centering
	\subfloat[\label{fig:wltp_exergyrate_EV}Electric Vehicle.]{\includegraphics[width=0.9\columnwidth]{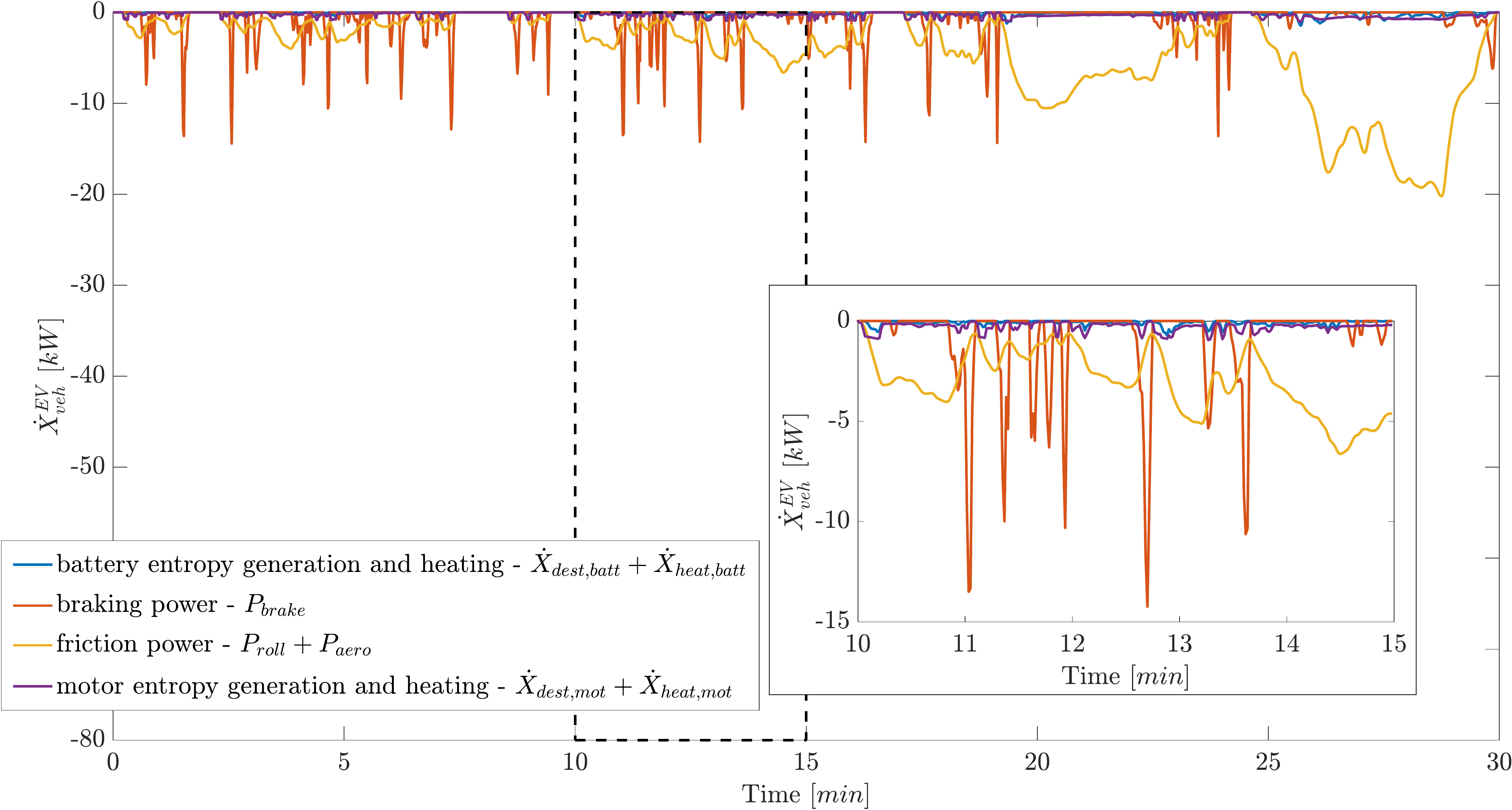}}\\
	\vspace{-1em}
	\subfloat[\label{fig:wltp_exergyrate_HEV}Hybrid Electric Vehicle.]{\includegraphics[width=0.9\columnwidth]{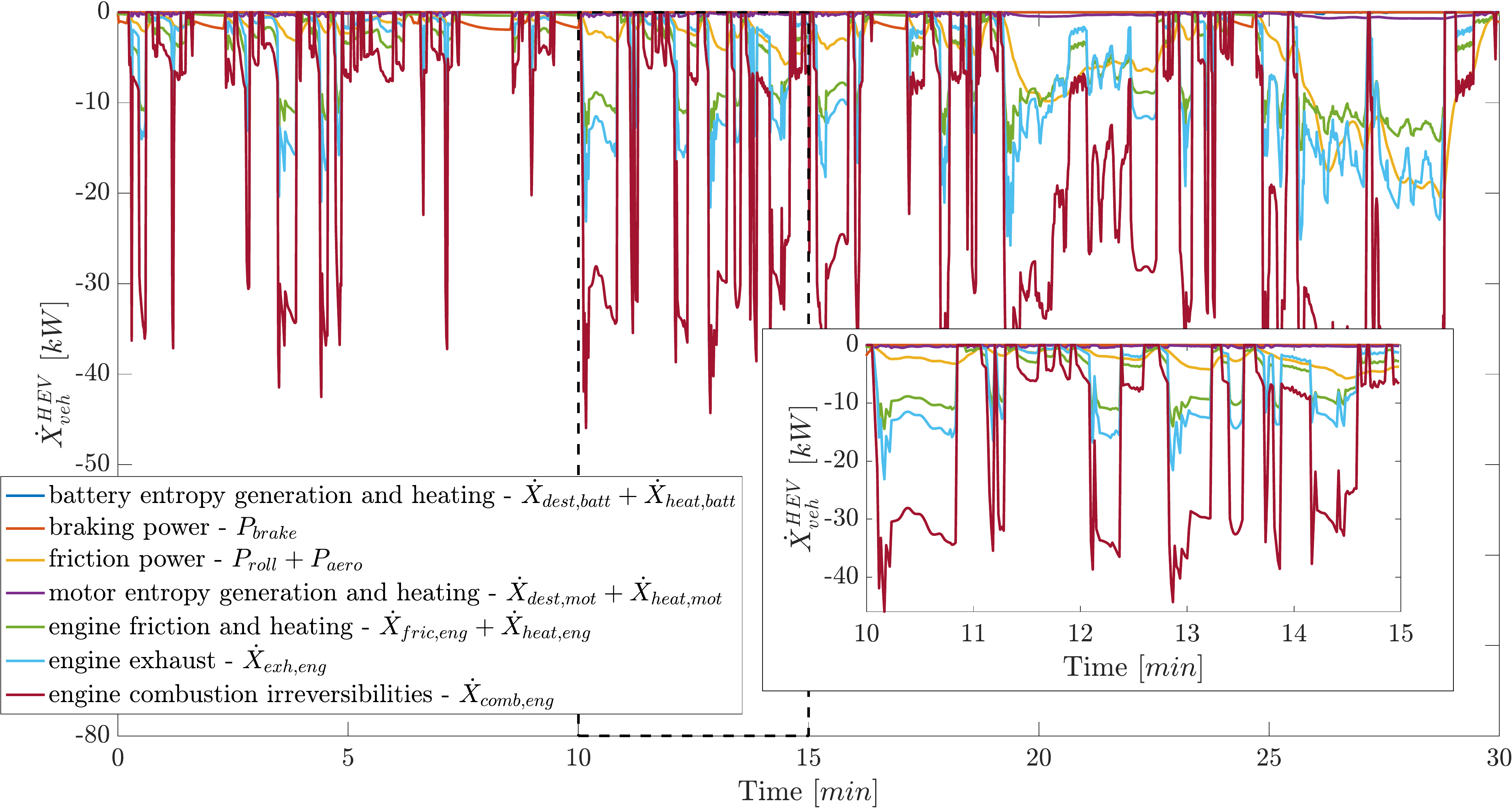}}
	\caption{\changeb{Exergy rate terms $\dot{X}$ and $P$ over the WLTP driving cycle. In the EV case study, the zoomed portion highlight that friction and braking terms are dominating the balance. In the HEV case, the engine related quantities are the principal source of availability loss.}\label{fig:wltp_exergyrate}}
\end{figure*}

\subsection{Results}
\label{section:simulation_results}
\changeb{For both the EV and HEV, the vehicle speed and the corresponding C-rate\footnote{The C-rate is computed dividing the battery current $I_{batt}$ by the cell nominal capacity $Q_{nom}^{cell}$.} profiles are shown in Fig. \ref{fig:wltp_cycle}.} In Fig. \ref{fig:wltp_cycle_hev}, the torque split between ICE and electric motor, computed by the energy management strategy, is shown. The management strategy, already implemented in the HEV simulator, is based on the Equivalent Consumption Minimization Strategy (ECMS) \cite{simona_book}. This is a Pontryagin's minimum principle-based algorithm that optimizes the power split between the battery pack and the ICE, \changeb{minimizing the fuel consumption.} %\change{As expected,} the engine is mostly used in the high speed \change{portion} of the cycle.

\changeb{To assess the exergetic behavior of the vehicle, the evolution of the exergy rate terms in $\dot{X}_{veh}^{EV}$ and $\dot{X}_{veh}^{HEV}$ is shown in Fig. \ref{fig:wltp_exergyrate}. The EV simulation results highlight that most of the available work is lost due to friction and braking (Fig. \ref{fig:wltp_exergyrate_EV}).} Moreover, a non-negligible portion of the losses is due to battery and motor heating, and entropy generation. In the HEV case, Fig. \ref{fig:wltp_exergyrate_HEV} shows that the drivetrain friction losses are overcame by the engine irreversibilities and exergy transfer to the environment (through \change{the exhaust gas}). In this context, the battery and motor losses are, in practice, negligible.

\begin{figure}[!tb]
	\centering
	\subfloat[\label{fig:ev_perc}Electric Vehicle.]{\includegraphics[width=\columnwidth]{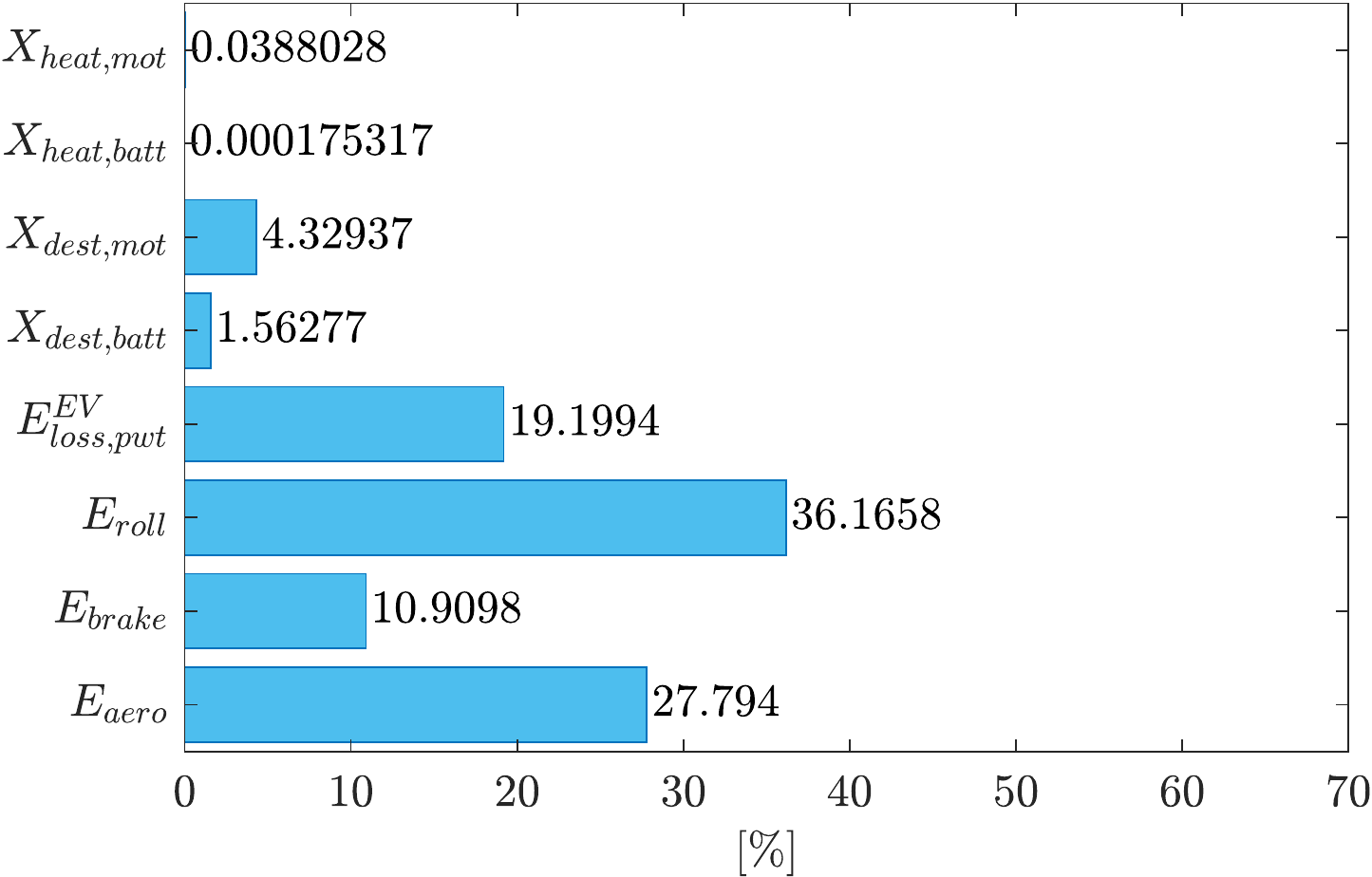}}\\
	\subfloat[\label{fig:hev_perc}Hybrid Electric Vehicle.]{\includegraphics[width=\columnwidth]{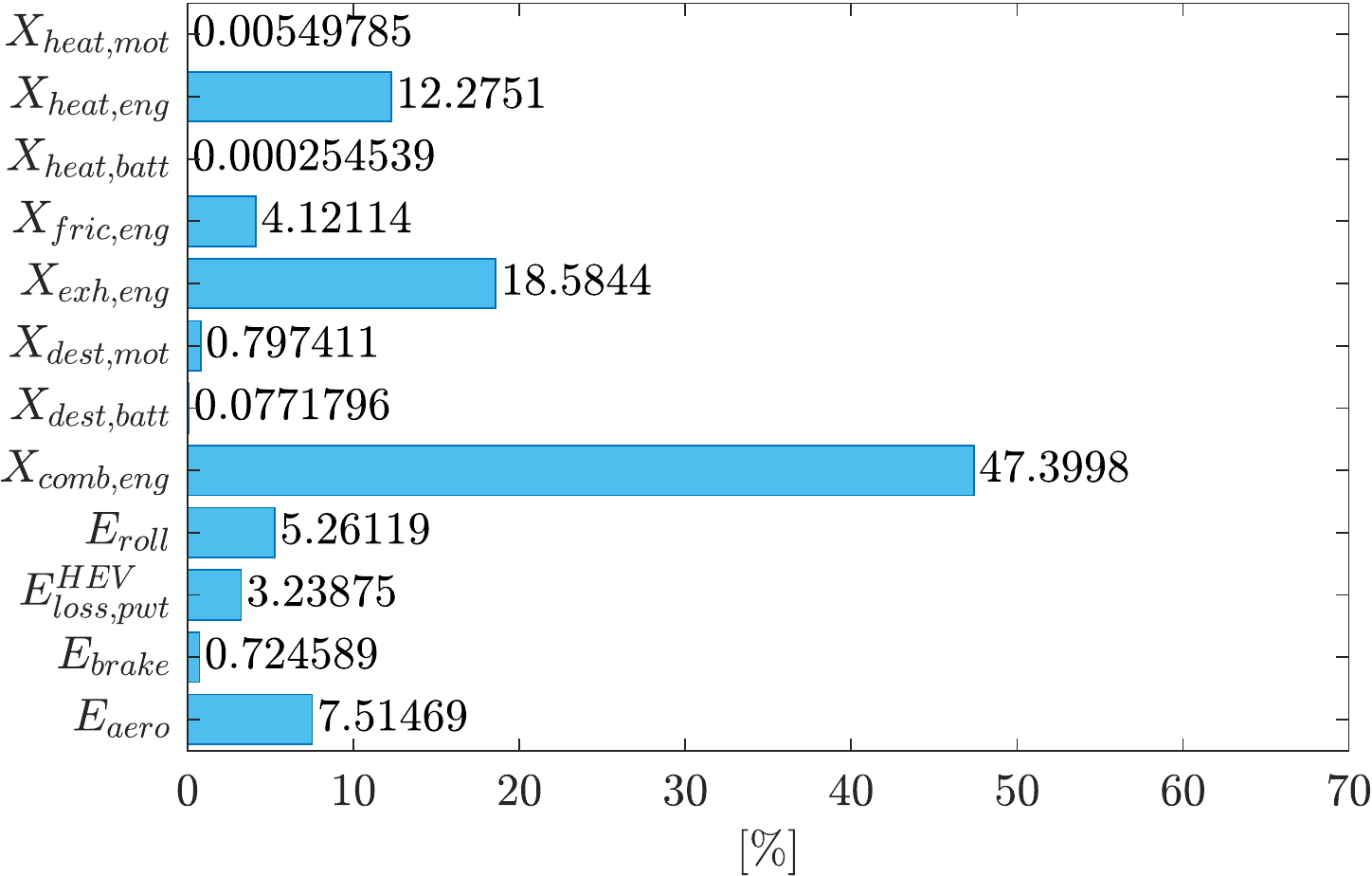}}
	\caption{\changeb{Vehicle exergy losses over the WLTP driving cycle. The contribution of each term is expressed as a percentage of the total losses  experienced along the driving cycle. In the EV case, rolling friction, aerodynamic drag, and $E_{loss,pwt}^{EV}$ are dominating. In the HEV scenario, the combustion irreversibilities term $X_{comb,eng}$ has the highest impact on the total balance.} }
	\label{fig:exergy_perc}
\end{figure}

\changeb{Integrating the exergy rate quantities in Fig. \ref{fig:wltp_exergyrate}, the exergy transfer and destruction terms are obtained\footnote{\changeb{The integration of the power terms $P$ of $\dot{X}_{long}$, defined as in Equation (\ref{eq:hamiltonian_exergy}), are energies expressed with the letter $E$.}}. The contribution of each term is expressed as a percentage of the total losses experienced along the driving cycle and computed as \resizebox{0.35\columnwidth}{!}{$X^{EV}_{veh}(t_f)- X^{EV}_{veh}(0)$} and \resizebox{0.39\columnwidth}{!}{$X^{HEV}_{veh}(t_f)- X^{HEV}_{veh}(0)$} for the EV and HEV, respectively.} In the EV case study (Fig. \ref{fig:ev_perc}), the electric motor exergy losses account for $5\%$ of the total, while the battery accounts for \mytilde$1\%$. \change{As shown in Fig. \changeb{\ref{fig:exergy_perc}a}, most of the losses are due to rolling friction, aerodynamic drag, and $E_{loss,pwt}^{EV}$. This is in line with the fact that efficiencies of energy storage/conversion devices in electric powertrains are generally around $90\%$ \cite{simona_book}. A key advantage of the proposed exergy-based modeling is the possibility to distinguish between the different sources of irreversibility, e.g., between \changeb{Joule losses in battery and electric motor.} This provides fundamental information to assess, \changeb{at vehicle and powertrain-leve}, how inefficiency is spread.} \changeb{In the HEV case study,} losses related to the battery and \change{electric motor} are almost negligible, being as small as the $1\%$ of the total \changeb{(see Fig. \ref{fig:hev_perc})}. The primary loss term is the \change{availability destruction} in the ICE due to \changeb{combustion reactions} (\mytilde$47\%$ of the total). $X_{comb,eng}$ is mainly related to the difference in chemical potential between the reactants and products participating in the combustion reaction. Moreover, \change{this term -- obtained from Equation (\ref{eq:eng_irrev}) --} lumps together the unmodeled exergy contributions given by blow-by gases, unburnt fuel, and intake air flow (overall, these terms account for \mytilde$5\%$ of the total balance \cite{razmara_exergy_engine}). The contribution of the combustion irreversibilities is also related to the fraction of fuel which can be converted into useful, \change{braking}, work: the higher the efficiency, the lower $X_{comb,eng}$. Considering the fuel consumption map in Fig. \ref{fig:ice_eff_map}, the ICE average efficiency is $28\%$ (computed along the driving cycle), \change{meaning that only a small portion of the fuel thermal energy is used to fulfill the traction power $P^{HEV}_{trac}$.} Together, the losses associated to the ICE account for more than $80\%$. This is expected as the ICE is rather an inefficient component in which, according to \cite{rakopoulos2009diesel} and depending on the operating conditions, only at most the $30$-$35\%$ of the fuel availability can be converted into \change{braking} work. 
\begin{figure}[!tb]
	\centering
	\subfloat[\label{fig:wltp_exergy_ev}Electric Vehicle.]{\includegraphics[width=\columnwidth]{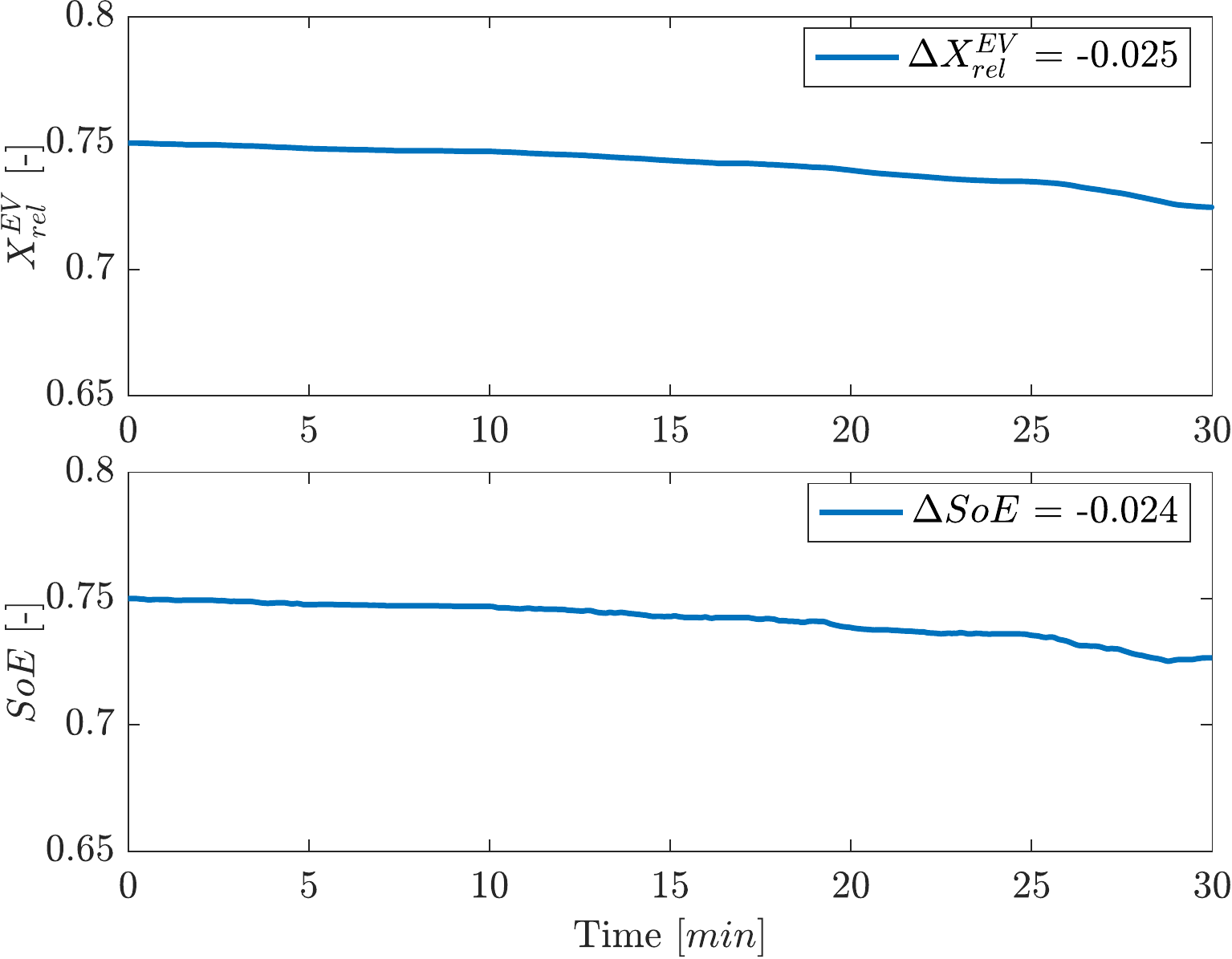}}
	\hfill
	\subfloat[\label{fig:wltp_exergy_hev}Hybrid Electric Vehicle.]{\includegraphics[width=\columnwidth]{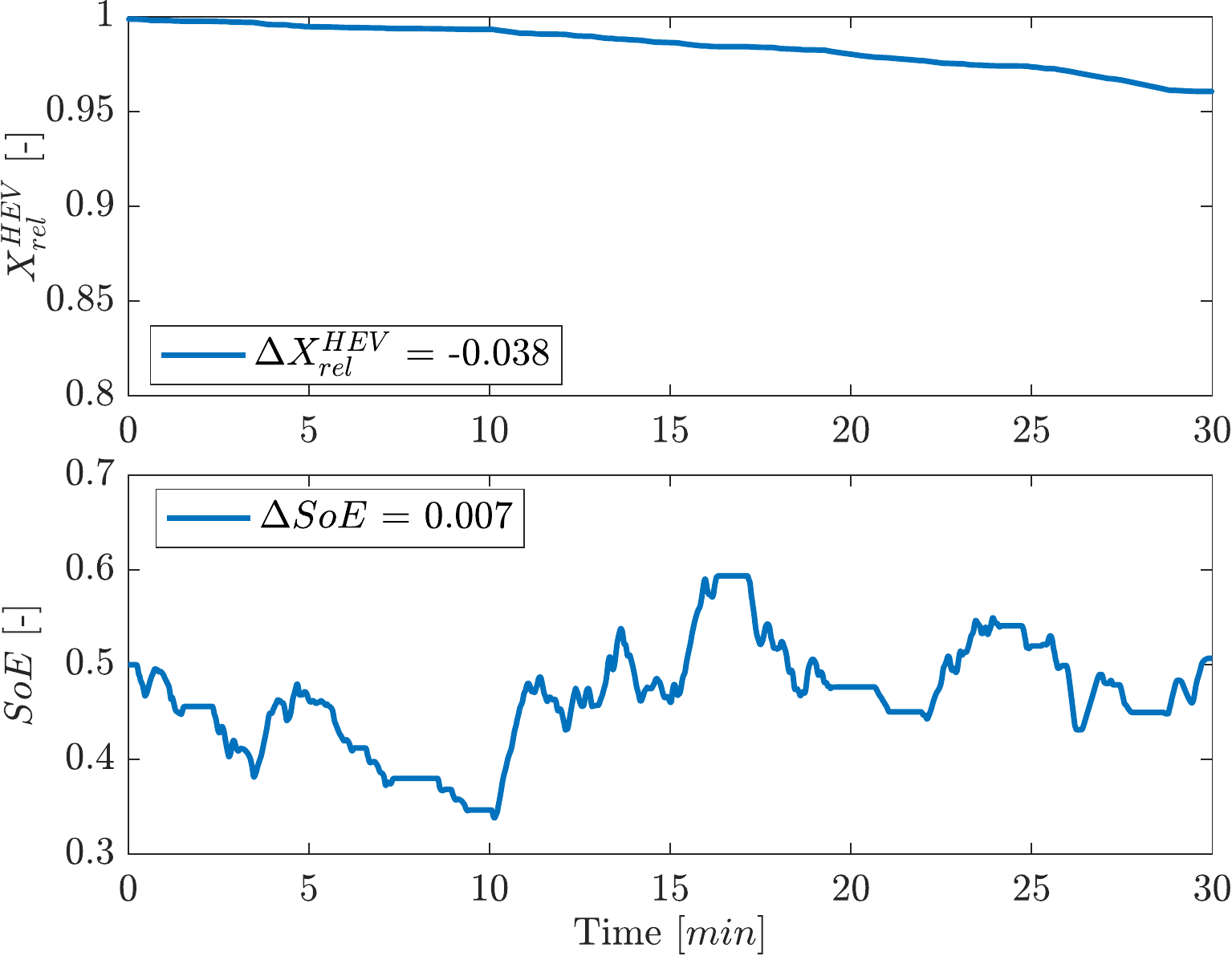}}
	\caption{\changeb{Relative exergies $X_{rel}^{EV}$ and $X_{rel}^{HEV}$ and battery $SoE$ profiles over the WLTP driving cycle.} \label{fig:wltp_exergy}}
\end{figure}

\changeb{In Fig. \ref{fig:wltp_exergy}, the relative exergy quantities $X_{rel}^{EV}$ and $X_{rel}^{HEV}$, defined in Equation (\ref{eq:ex_normalization}), are computed and compared to the $SoE$. For what concerns the EV (Fig. \ref{fig:wltp_exergy_ev}), at the beginning of the driving cycle the battery pack is charged to $SoE=0.75$. The evolution of the $SoE$ and $X_{rel}^{EV}$ is similar: this is expected since the battery is the only storage device and its charged capacity defines the maximum availability of the system. Computing the difference between the initial (at $0s$) and final (at $t_f$) state of $SoE$ and $X_{rel}^{EV}$ leads to $\Delta SoE = -0.024$ and $\Delta X_{rel}^{EV} = -0.025$, respectively. The lower value of $\Delta X_{rel}^{EV}$ with respect to $\Delta SoE$ is due to the exergy balance formulation which, according to Equation (\ref{eq:exergy_HEV}), takes into account also the losses due to entropy generation and heat transfer. The discrepancy between $\Delta X_{rel}^{EV}$ and $\Delta SoE$ proves the ability of the exergy-based modeling in quantifying the availability loss not just as a function of the electrical work $P_{batt}\cdot I_{batt}$ but also of the interaction with the surroundings (heat transfer) and entropy generation (e.g., Joule losses). In the HEV case study, the energy management strategy keeps the battery $SoE$ around a reference value of $0.5$ during the whole driving cycle (see Fig. \ref{fig:wltp_exergy_hev}, bottom plot). Thus, the decrease in the relative exergy term $X_{rel}^{HEV}$ is due to the fuel consumed by the ICE and converted into mechanical work or lost because of the engine irreversibilities, friction, heat transfer, and exhaust gas. For the WLTP driving cycle, a value of $\Delta X_{rel}^{HEV}$ equal to $-0.038$ corresponds to a fuel consumption of $1.3kg$.}

\section{Conclusion}
\label{section:conclusion}
In this paper, a comprehensive exergy-based modeling framework for ground vehicles is proposed. \changeb{Starting from the formulation of the exergy balance equations for the vehicle's longitudinal dynamics and its powertrain components, namely, electrochemical energy storage device, electric motor, and ICE, the framework is applied to two case studies: an EV and a HEV. The analysis allows to quantify, locate, and rank the sources of losses.} In the EV case, the exergy balance shows that most of the energy stored in the battery is used to fulfill the traction power $P^{EV}_{trac}$, needed for the vehicle motion. The principal sources of availability loss are the rolling friction and aerodynamic drag, with the battery and electric motor contributing for the 1\% and 5\% of the losses, respectively. In the HEV case study, 80\% of the exergy losses are due to the ICE. In particular, irreversibilities related to the combustion process have the highest impact on the total balance (\mytilde 47\%).

\changeb{The development of the proposed modeling framework is the first step for the design of management strategies aimed at minimizing the ground vehicle exergy losses rather than its fuel consumption.}

\section*{Acknowledgments}
Unclassified. DISTRIBUTION STATEMENT A. Approved for public release; distribution is unlimited. Reference herein to any specific commercial company, product, process, or service by trade name, trademark, manufacturer, or otherwise does not necessarily constitute or imply its endorsement, recommendation, or favoring by the United States Government or the Dept. of the Army (DoA). The opinions of the authors expressed herein do not necessarily state or reflect those of the United States Government or the DoD, and shall not be used for advertising or product endorsement purposes.

%% TABLE PARAMS
\renewcommand{\arraystretch}{1.14}
\begin{table*}[!htb]
\centering
\resizebox{1\columnwidth}{!}{
	\begin{tabular}{| *{5}{c|} }
		\hline 
		\textbf{Parameter} & \textbf{Description} & \textbf{EV} & \textbf{HEV} & \textbf{Unit} \\
		\hline\hline
	    $\rho_{air}$ & Air density & \multicolumn{2}{c|}{$1.18$} & $kg/m^3$\\
	    	\hline
		$g$ & Gravitational acceleration &\multicolumn{2}{c|}{$9.81$} & $m/s^2$ \\
		\hline
		$A_f$ & Vehicle frontal area &$2.34$ \scriptsize{\cite{sherman2016drag}} & \GP{$2.21$} \scriptsize{\cite{buggaveeti2017longitudinal}} & $m^2$\\
		\hline
		$C_d$ & Aerodynamic drag coefficient &$0.24$ \scriptsize{\cite{sherman2016drag}} & \GP{$0.25$} \scriptsize{\cite{buggaveeti2017longitudinal}} & $-$ \\
		\hline
		$\eta_{diff}$ & Efficiency of the differential &\multicolumn{2}{c|}{$0.98$}& $-$ \\
		\hline
		$k_{roll}$ & Rolling friction coefficient &\multicolumn{2}{c|}{$0.009$} & $-$ \\
		\hline
		$m_{veh}$ & Vehicle mass & \GP{$2108$} \scriptsize{\cite{grunditz2016performance}} & \GP{$1360$} \scriptsize{\cite{cheng2011specifications}} & $kg$\\
		\hline
		$\mathcal{R}_{wh}$ & Wheel radius &$0.483$ \scriptsize{\cite{grunditz2016performance}}& \GP{$0.3$} \scriptsize{\cite{cheng2011specifications}}  & $m$\\
		\hline
		$E_{nom}$ & Battery pack nominal energy & \textcolor{black}{$90$} \scriptsize{\cite{grunditz2016performance}} & \textcolor{black}{$1$} \scriptsize{\cite{cheng2011specifications}} & $kWh$ \\
		\hline
		$V_{nom}$ & Battery pack nominal voltage & \textcolor{black}{$400$} \scriptsize{\cite{grunditz2016performance}} & \GP{$201.6$} \scriptsize{\cite{cheng2011specifications}} & $V$ \\
		\hline
		$N_s$ & Battery pack series cells configuration & \textcolor{black}{$110$}* & $55$* & $-$ \\
		\hline
		$N_p$ & Battery pack parallel cells configuration & \textcolor{black}{$46$}* & $1$* & $-$ \\
		\hline
		$\mathrm{C}_{cell}$ & \changeb{Battery cell} thermal capacity &\multicolumn{2}{c|}{$156.3790$**}& $J/K$\\
		\hline
		$\mathrm{h}_{out,cell}$& \changeb{Battery cell} thermal transfer coefficient & \multicolumn{2}{c|}{$0.2085$**}& $W/K$\\
		\hline
		$\tau_{mot,max}$ & Maximum electric motor torque (the EV is equipped with 2 motors) & $2\cdot329$ \scriptsize{\cite{teslamot_max}} & $200$ & $Nm$\\ 
		\hline
		$P_{mot,max}$ & Maximum electric motor power (the EV is equipped with 2 motors) &$2\cdot193$ \scriptsize{\cite{teslamot_max}}  & $125.6$ & $kW$\\
		\hline
		$\mathrm{C}_{mot,iron}$ & Iron thermal capacity &\multicolumn{2}{c|}{$33401$ \scriptsize{\cite{rajput_electric_motor}}}& $J/K$\\
		\hline
		$\mathrm{C}_{mot,copper}$ & Copper thermal capacity &\multicolumn{2}{c|}{$4903.6$ \scriptsize{\cite{rajput_electric_motor}}}& $J/K$\\
		\hline
		$\mathrm{h}_{mot,iron}$ & Iron thermal transfer coefficient &\multicolumn{2}{c|}{$ 66.6667$ \scriptsize{\cite{rajput_electric_motor}}}& $W/K$\\
		\hline
		$\mathrm{h}_{mot,copper}$ & Copper thermal transfer coefficient &\multicolumn{2}{c|}{$   27.0270$ \scriptsize{\cite{rajput_electric_motor}}}& $W/K$\\
		\hline
		$k_h$ & Electric motor iron losses coefficient &\multicolumn{2}{c|}{$27.543$ \scriptsize{\cite{rajput_electric_motor}}}& $rad\cdot W\cdot J/A^2$\\
		\hline
		$k_f$ & Electric motor friction losses coefficient &\multicolumn{2}{c|}{$10^{-3}$ \scriptsize{\cite{rajput_electric_motor}}}& $W/\left(rad\cdot s\right)^2$\\
		\hline
		$\lambda_{pm}$ & Electric motor permanent magnet flux linkage &\multicolumn{2}{c|}{$0.1194$}& $Wb$\\
		\hline
		$N_{pp}$ & Electric motor pole pairs &\multicolumn{2}{c|}{$4$}& $-$\\
		\hline
		$L_q$ & Electric motor $q$-axis inductance &\multicolumn{2}{c|}{$4.1840\cdot10^{-1}$} & $mH$\\
		\hline
		$L_d$ & Electric motor $d$-axis inductance &\multicolumn{2}{c|}{$3.752\cdot10^{-1}$}& $mH$\\
		\hline
		$R_{s,0}$ & Electric motor resistance at $T_0$ &\multicolumn{2}{c|}{$4.7973$ \scriptsize{\cite{rajput_electric_motor}}}& $\Omega$\\
		\hline
		$\xi$ & Electric motor resistance coefficient &\multicolumn{2}{c|}{$0.0039$ \scriptsize{\cite{rajput_electric_motor}}}& $\Omega/K$\\
		\hline
		$\mathcal{R}_{gas}$ & Ideal gas constant &$-$ & $8.31$  \scriptsize{\cite{heywood_book}} & $J/(K\cdot mol)$\\
		\hline
		$LHV$ & Fuel lower heating value &$-$ & $47.3$ & $MJ/kg$\\
		\hline
		$AFR_{stoich}$ & Stoichiometric air-fuel ratio &$-$ & $14.6$ & $-$\\
	     \hline
		$B$ & Cylinder bore & $-$ & $0.0805$ \scriptsize{\cite{prius_ts}}& $m$\\
		\hline
		$\mathcal{V}_d$ & ICE displacement & $-$ & $1.8$ \scriptsize{\cite{prius_ts}}& $l$\\
		\hline
		$\mathcal{V}_{tank}$ & Fuel tank volume & $-$ & $43$  \scriptsize{\cite{prius_ts}}& $l$\\
		\hline
		$\rho_{fuel}$ & Fuel density & $-$ & $755$ \scriptsize{\cite{guzzella_book}}& $kg/m^3$\\
		\hline
		$a_{comb}$ & Combustion reaction $CO_2$ coefficient & $-$ & $8$ &$-$  \\
		\hline
		$b_{comb}$ & Combustion reaction $H_2O$ coefficient & $-$ & $9$ &$-$  \\
		\hline
		$c_{comb}$ & Combustion reaction $N_2$ coefficient & $-$ & $47$ &$-$  \\
		\hline
		$f_{N_2,0}$ & Reference \changeb{state} $N_2$ molar fraction & $-$ & $0.7567$ \scriptsize{\cite{mahabadipour_engine}} & $-$ \\
		\hline
		$f_{O_2,0}$ & Reference \changeb{state} $O_2$ molar fraction & $-$ & $0.2035$ \scriptsize{\cite{mahabadipour_engine}}& $-$ \\
		\hline
		$f_{CO_2,0}$ & Reference \changeb{state} $CO_2$ molar fraction & $-$ & $0.0003$ \scriptsize{\cite{mahabadipour_engine}}&$-$  \\
		\hline
		$f_{H_2O,0}$ & Reference \changeb{state} $H_2O$ molar fraction & $-$ & $0.0303$ \scriptsize{\cite{mahabadipour_engine}}& $-$ \\
		\hline	
		$f_{others,0}$ & \makecell{Reference \changeb{state} molar fraction \\ of others species}  & $-$ & $0.0092$ \scriptsize{\cite{mahabadipour_engine}}& $-$ \\
		\hline		
		$T_{eng}$ &  Mixture temperature & $-$ & $677.23$ \scriptsize{\cite{heywood_book}}& $K$ \\
		\hline	
		$T_{c}$ &  Coolant temperature & $-$ & $373.15$ \scriptsize{\cite{heywood_book}}& $K$ \\
		\hline	
		$k_g$ &  Mixture conductivity & $-$ & $0.05$ \scriptsize{\cite{heywood_book}}& $W/(m\cdot K)$ \\
		\hline	
		$\mu_g$ &  Mixture viscosity & $-$ & $3.26\cdot 10^{-5}$ \scriptsize{\cite{heywood_book}}& $kg/(s\cdot m)$ \\
		\hline	
		$b$ & Coefficient for Taylor\&Toong correlation & $-$ & $0.75$ \scriptsize{\cite{heywood_book}}& $-$ \\
		\hline	
	\end{tabular}}
		\begin{flushleft}
		\scriptsize{\textit{*: the series/parallel configuration of the battery pack is obtained combining NMC cylindrical cells (with nominal capacity $Q_{nom}^{cell} = 4.85Ah$), available at the Stanford Energy Control Lab, to meet the target energy ($E_{nom}$) and voltage ($V_{nom}$) specifications.}}\\
		\scriptsize{\textit{**: identified from experimental data available at the Stanford University Energy Control Lab.}}
	\end{flushleft}\vspace{-0.5em}
	\caption{Vehicle parameters. Values without an explicit reference are obtained from \citep{powertrainblockset}.}
	\label{tab:vehicle_parameters_table}
\end{table*}

\begin{nomenclature}
\scriptsize{
\begin{deflist}[AAAAA] %[AAAA] if you have 4 letters max for example
\defitem{$\alpha,\beta$}\defterm{Coefficients for the electric motor thermal properties $[\mathrm{-}]$}
\defitem{$\eta$}\defterm{Efficiency $[\mathrm{-}]$}
\defitem{$a,b$}\defterm{Coefficients for Taylor\&Toong correlation $[\mathrm{-}]$}
\defitem{$C_d$}\defterm{Aerodynamic drag coefficient $[-]$}
\defitem{$f$}\defterm{Molar fraction $[-]$}
\defitem{$k_{roll}$}\defterm{Rolling friction coefficient $[-]$}
\defitem{$N_s,N_p$}\defterm{Battery series/parallel configuration $[\mathrm{-}]$}
\defitem{$N_{pp}$}\defterm{Electric motor pole pairs $[\mathrm{-}]$}
\defitem{$SoC$}\defterm{State of Charge $[\mathrm{-}]$}
\defitem{$SoE$}\defterm{State of Energy $[\mathrm{-}]$}
\defitem{$t$}\defterm{Time $[s]$}
\defitem{$t_f$}\defterm{Driving cycle duration $[s]$}
\defitem{$B$}\defterm{Cylinder bore $[m]$}
\defitem{$\mathcal{R}_{wh}$}\defterm{Wheel radius $[m]$}
\defitem{$A_f$}\defterm{Vehicle frontal area $[m^2]$}
\defitem{$g$}\defterm{Gravitational acceleration $[m/s^2]$}
\defitem{$v,\dot{v}$}\defterm{Speed and acceleration $[m/s],[m/s^2]$}
\defitem{$\mathcal{V},\dot{\mathcal{V}}$}\defterm{Volume variation $[m^3],[m^3/s]$}
\defitem{$AFR_{stoich}$}\defterm{Air-fuel ratio at stoichiometric $[mol]$}
\defitem{$n,\dot{n}$}\defterm{Moles and molar flow rate  $[mol],[mol/s]$}
\defitem{$m,\dot{m}$}\defterm{Mass and mass flow rate $[kg],[kg/s]$}
\defitem{$M$}\defterm{Molar mass $[kg/mol]$}
\defitem{$\rho$}\defterm{Density $[kg/m^3]$}
\defitem{$T$}\defterm{Temperature $[K]$}
\defitem{$E,\dot{P}$}\defterm{Energy and power $[J],[W]$}
\defitem{$Q,\dot{Q}$}\defterm{Heat transfer and heat transfer rate $[J],[W]$}
\defitem{$W,\dot{W}$}\defterm{Work and work rate $[J],[W]$}
\defitem{$X,\dot{X}$}\defterm{Exergy and exergy rate $[J],[W]$}
\defitem{$S,\dot{S}$}\defterm{Entropy and entropy rate $[J],[W]$}
\defitem{$\mathrm{C}$}\defterm{Thermal capacity $[J/K]$}
\defitem{$\mathrm{h}$}\defterm{Thermal transfer coefficient $[W/K]$}
\defitem{$\mathcal{R}_{gas}$}\defterm{Ideal gas constant $[J/(mol\cdot K)]$}
\defitem{$LHV$}\defterm{Fuel lower heating value $[MJ/kg]$}
\defitem{$\psi$}\defterm{Exergy flux $[J/mol]$}
\defitem{$h$}\defterm{Specific enthalpy $[J/mol]$}
\defitem{$s$}\defterm{Specific entropy $[J/(mol\cdot K)]$}
\defitem{$I$}\defterm{Current $[A]$}
\defitem{$V$}\defterm{Voltage $[V]$}
\defitem{$R$}\defterm{Resistance $[\Omega]$}
\defitem{$\xi$}\defterm{Electric motor resistance coefficient $[\Omega/K]$}
\defitem{$L_q,L_d$}\defterm{Electric motor $q$ and $d$ axes inductances $[mH]$}
\defitem{$\lambda_{pm}$}\defterm{Permanent magnet flux linkage $[Wb]$}
\defitem{$k_h$}\defterm{Electric motor iron losses coefficient $[rad\cdot W\cdot J/A^2]$}
\defitem{$k_f$}\defterm{Electric motor friction losses coefficient $[W/\left(rad\cdot s\right)^2]$}
\defitem{$F$}\defterm{Force $[N]$}
\defitem{$\tau$}\defterm{Torque $[Nm]$}
\defitem{$\omega$}\defterm{Rotational speed $[rad/s]$}
\end{deflist}}
\end{nomenclature}

\begin{notation}
\scriptsize{
\begin{deflist}[AAA] %[AAAA] if you have 4 letters max for example
\defitem{$j,k$}\defterm{Chemical species in the intake and exhaust manifolds}
\defitem{$\mathcal{K}$}\defterm{Set collecting the chemical species $\{N_2,O_2,H_2O,CO_2\}$}
\defitem{$d\mathcal{X}$}\defterm{Differential of a variable $\mathcal{X}$}
\defitem{$\Delta\mathcal{X}$}\defterm{Variation of a variable $\mathcal{X}$ over the driving cycle: $\mathcal{X}(0)-\mathcal{X}(t_f)$}
\defitem{$\dot{\mathcal{X}}$}\defterm{Time derivative of a variable $\mathcal{X}$}
\defitem{$\mathcal{X}_0$}\defterm{Variable $\mathcal{X}$ at the reference state}
\defitem{$\mathcal{X}^\star$}\defterm{Variable $\mathcal{X}$ at the restricted state}
\defitem{$\mathrm{min}$}\defterm{Minimum function}
\defitem{$\mathrm{sign}$}\defterm{Sign function}
\end{deflist}}
\end{notation}

\begin{abbrv}
\scriptsize{
\begin{deflist}[AAAAA] %[AAAA] if you have 4 letters max for example
\defitem{$aero$}\defterm{Aerodynamic}
\defitem{$batt$}\defterm{Battery pack}
\defitem{$ch$}\defterm{Chemical}
\defitem{$comb$}\defterm{Combustion}
\defitem{$d$}\defterm{Drag}
\defitem{$Dest,dest$}\defterm{Destruction}
\defitem{$diff$}\defterm{Differential}
\defitem{$eng$}\defterm{Engine}
\defitem{$exh$}\defterm{Exhaust}
\defitem{$fric$}\defterm{Friction}
\defitem{$gen$}\defterm{Generation}
\defitem{$heat$}\defterm{Heat transfer}
\defitem{$intk$}\defterm{Intake}
\defitem{$long$}\defterm{Longitudinal}
\defitem{$max$}\defterm{Maximum}
\defitem{$mot$}\defterm{Motor}
\defitem{$nom$}\defterm{Nominal}
\defitem{$oc$}\defterm{Open circuit}
\defitem{$ph$}\defterm{Physical}
\defitem{$pm$}\defterm{Permanent magnet}
\defitem{$pp$}\defterm{Pole pairs}
\defitem{$pwt$}\defterm{Powertrain}
\defitem{$ref$}\defterm{Reference}
\defitem{$rel$}\defterm{Relative}
\defitem{$roll$}\defterm{Rolling}
\defitem{$spec$}\defterm{Specific}
\defitem{$stoich$}\defterm{Stoichiometric}
\defitem{$Surr$}\defterm{Surroundings}
\defitem{$tot$}\defterm{Total}
\defitem{$trac$}\defterm{Traction}
\defitem{$veh$}\defterm{Vehicle}
\defitem{$wh$}\defterm{Wheel}
\end{deflist}}
\end{abbrv}

\bibliographystyle{elsarticle-num}
\bibliography{ref}

\end{document}